\documentclass[12pt,preprint]{aastex}

\usepackage{amsmath}
\usepackage{amssymb}
\usepackage{amsmath}
\usepackage{mathrsfs}
\usepackage{natbib}

\shorttitle{Jets, Disks and Field Topology}
\shortauthors{Beckwith, Hawley, Krolik}

\begin{document}

\title{The Influence of Magnetic Field Geometry on the Evolution
of Black Hole Accretion Flows: Similar Disks, Drastically Different Jets}

\author{Kris Beckwith and John F. Hawley}
\affil{Astronomy Department\\
University of Virginia\\ 
P.O. Box 400325\\
Charlottesville, VA 22904-4325}

\email{krb3u@virginia.edu; jh8h@virginia.edu}

\and
\author{Julian H. Krolik}
\affil{Department of Physics and Astronomy\\
Johns Hopkins University\\
Baltimore, MD 21218}

\email{jhk@pha.jhu.edu}

\begin{abstract}

Because the magnetorotational instability is capable of exponentially
amplifying weak preexisting magnetic fields, it might be hoped that the
character of the magnetic field in accretion disks would be independent
of the nature of the seed field.  However, the divergence-free
nature of magnetic fields in highly conducting fluids ensures that
their large-scale topology is preserved, no matter how greatly the
field intensity is changed.  By performing global two-dimensional
and three-dimensional general relativistic magnetohydrodynamic disk
simulations with several different topologies for the initial magnetic
field, we explore the degree to which the character of the flows around
black holes depends on the initial topology.  We find that while the
qualitative properties of the accretion flow are nearly independent
of field topology, jet-launching is very sensitive to it: a sense of
vertical field consistent for at least an inner disk inflow time is
essential to the support of strong jets.

\end{abstract}


\section{Introduction}\label{intro}

 Magnetic fields are now generally acknowledged to be essential
to accretion \citep{Balbus:1998}.  Amplified exponentially from weak seed
fields by the magneto-rotational instability (MRI), they are stretched
in a consistent direction so as to produce a net stress $\langle -B_r
B_\phi\rangle/4\pi$ capable of carrying enough angular momentum outward
through the disk to support sizable accretion rates.   Because the
nonlinear state of the MRI is fully-developed magnetohydrodynamic (MHD)
turbulence, it is often thought (or at least hoped) that its properties
have little to do with the character of the seed magnetic field brought
to the disk with the accretion flow.

 However, there are certain properties of the magnetic field
that are preserved despite the development of nonlinear turbulence.
Because magnetic fields are divergence-free and the plasma in accretion
flows has very little electrical resistivity, the global topology
of the field persists, no matter what happens.  It is therefore
natural to ask whether any interesting properties of the accretion
flow may depend on the global field topology.  For example, could
the field topology alter the relation between surface density and
accretion rate?   Or how the stress varies through the marginally
stable region?

  In addition to swallowing matter, accreting black holes often
expel relativistic jets.  Although the mechanisms by which these jets are
launched have long been the subject of investigation and speculation,
the current consensus focuses on a combination of large-scale magnetic
fields and the rotation of either the central hole, the accretion
disk, or both.  Two specific MHD models have received the greatest
attention: the Blandford-Znajek mechanism \citep{Blandford:1977}
and the Blandford-Payne mechanism \citep{Blandford:1982}.  In the
Blandford-Payne model a large-scale vertical magnetic field is anchored
in the disk, rotating with the orbital frequency.  Above and below the
disk magnetic tension dominates, and the field lines force the matter to
rotate with this same frequency.  If the fieldlines are angled outward
sufficiently with respect to the disk, there can be a net outward force,
accelerating the matter along the rotating fieldlines.
The Blandford-Znajek model also depends upon rotating field lines, but
the source of that rotation (and the source of the power in the jet)
is the black hole rather than the disk: any field lines connecting the
immediate vicinity of the black hole to infinity are forced to rotate
by frame-dragging, so that Poynting flux carries off the energy.

 Given the strong dependence of both the Blandford-Znajek and the
Blandford-Payne mechanisms on the nature of the magnetic field in the
disk and near the black hole, it is equally natural to inquire whether
the topology of the field embedded in the accretion flow has any influence
on their operation.  

 It is the goal of this paper to begin to investigate these questions,
both as they relate to accretion disk dynamics and as they relate to
jet properties.  Unfortunately, it is difficult even to scratch the
surface by purely analytic techniques \cite[although there have been some
attempts using highly-simplified models:][]{Krolik:1999a,Gammie:1999}.
On the other hand, direct numerical simulations provide a means by which
aspects of these problems can be investigated in detail.  \cite{HK:2002}
made a first exploration of the dependence of accretion properties on
field topology in a highly-simplified approximation: Newtonian dynamics
in a pseudo-Newtonian potential.   They found that a purely toroidal
initial field led to a systematically smaller accretion rate than one
whose initial field was dipolar.

More recently, fully general relativistic MHD  (GRMHD) simulations have
become possible \citep{McKinney:2004, De-Villiers:2005,Hawley:2006}.
These works have focused primarily on a single initial field structure:
a set of nested dipole field loops contained entirely within the
initial gas torus.  Other field geometries have been examined, but
only in rather limited fashion.  For example, \cite{De-Villiers:2005}
simulated the evolution of a torus with an initial toroidal field around
a Schwarzschild black hole and noted the lack of any resulting jet,
but did not consider why no jet formed.  \cite{McKinney:2004}, using
axisymmetric simulations, compared the electromagnetic luminosity
of the standard dipole jet with that resulting from a few alternate
magnetic field topologies.  In particular, they studied cases in
which the initial field was arranged in loops of alternating sign
arranged either vertically or horizontally within the initial torus.
The principal effect they reported was a factor of 2--3 reduction
in the ratio of the time-averaged jet Poynting flux to the mean mass
accretion rate.  They also noted that an initial field that is purely
vertical led to significant enhancements in both the mean mass accretion
rate and the time-averaged jet Poynting flux.  Axisymmetric simulations
of tori threaded by net vertical field led \cite{De-Villiers:2006} to
concur that such a field can greatly enhance jet power.  On the basis
of these results, \cite{McKinney:2007} proposed that black holes
inevitably become threaded by an organized vertical field if the initial
disk contains a large-scale poloidal field with few parity changes.
The intriguing character of these results motivates a more systematic
study, one that employs three-dimensional simulations, considers the
accretion flow as well as the jet, and investigates time-dependence.

The question immediately arises: how can one best attempt to describe the
very complex mixture of topologies likely to be found in real accretion
flows?  The magnetic field near a black hole can be thought of as simply
what the accretion process itself has self-consistently brought there
from whatever source supplies it.   In order to gain understanding of
what, in realistic circumstances, must be a truly complex combination of
structures, in this paper we do not try to construct a single ``realistic"
field geometry; instead, we study the effects due to individual members
of a ``basis set" of field topologies.  Our hope is that by understanding
the action of pure types, we can gain some understanding of how realistic
mixtures behave.

To be specific, we will begin with a carefully chosen set of different
kinds of closed field loops that we expect to be generic.  Closed field
loops have the advantage of requiring many fewer free parameters
to describe them fully within a simulation than do field geometries
with net flux.  Net flux geometries may well exist in Nature and have
important effects; we will consider them in the second stage of this
project.   We begin with the conjecture that, at a qualitative level,
the different sorts of closed field loop behavior can all be represented
by a combination of those due to: dipolar loops, or fieldlines initially
entirely poloidal and which cross the equatorial plane at well-separated
radii; toroidal loops, which encircle the black hole; and quadrupolar
loops, which (again, in the initial state) are poloidal, but close
without crossing the equatorial plane.  For each of these cases, we have
run a lengthy three-dimensional GRMHD simulation whose results we hope
characterize the accretion flow structure created by that field topology.

We have also run a series of high resolution axisymmetric simulations of
each of the initially poloidal topologies (i.e. dipole and quadrupole). In
addition, in order to understand better the implications of the finite
lifetimes of closed loops, we have performed an axisymmetric simulation
whose initial magnetic field structure was a series of dipolar loops,
each contained within a span of radii narrow enough that the difference
in inflow times across a loop is comparable to the equilibration time
for the magnetic field in the jet.

The rest of this work is structured as follows. In \S\ref{numerics}
we briefly review the numerical scheme employed to solve the
equations of GRMHD and give a detailed description of the initial
conditions used to generate each topology. In \S\ref{flowstructure}
we examine the late time accretion flow structure that evolves from
each of the initial field topologies, paying attention to each of the
structures (main disk body, magnetized corona) that were identified by
\cite{De-Villiers:2003b} as arising from the evolution of the dipole
field topology.  In \S\ref{jets} we examine the properties of the third
structure identified by \cite{De-Villiers:2003b}, the unbound outflow,
and examine how the initial evolution of the field during the linear
growth phase of the MRI gives rise to the components of the accretion
flow at late times. Finally in \S\ref{conclusion} we summarize our
results and highlight their importance for jets in astrophysical settings.

\section{Numerical Details}\label{numerics}

In this work we undertake a study of both two- and three-dimensional
simulations to investigate the influence of magnetic field topology
on both the evolved accretion flow structure and relativistic jet
formation.  In doing so we continue a program of black hole accretion
disk simulations begun in \cite{De-Villiers:2003b}, \cite{Hirose:2004},
\cite{De-Villiers:2005}, \cite{Krolik:2005} and \cite{Hawley:2006}.
Three-dimensional simulations, such as those presented in the previous
work, are the only means by which one can gain insight into the
evolved accretion flow properties due to the fundamental restriction posed by
the anti-dynamo theorem \cite[]{Moffatt:1978}.  Further, toroidal field
models cannot be studied in axisymmetry, since the toroidal MRI requires
nonzero azimuthal wavevectors.  On the other hand, because jet formation
occurs promptly
during the evolution of the flow \cite[]{Hawley:2006}, restrictions due
to the anti-dynamo theorem are less critical for jet studies.  In this
case the higher resolution afforded by the two-dimensional simulations
is particularly valuable.

The simulation code we use is
described in \cite{De-Villiers:2003}.  This code solves the equations of
ideal non-radiative MHD in the static Kerr metric of a rotating black hole
using Boyer-Lindquist coordinates.  Values are expressed in gravitational
units $(G = M = c = 1)$ with line element $ds^{2} = g_{tt} dt^{2} +
2g_{t \phi} dt d\phi + g_{rr} dr^{2} + g_{\theta \theta} d \theta^{2}
+ g_{\phi \phi} d\phi^{2}$ and signature $(-,+,+,+)$.  Since we are
focusing in this paper on the influence of the magnetic field topology,
we fix the black hole spin to $a/M =0.9$ for all of the new simulations
reported here.

The relativistic fluid at each grid zone is described by its density
$\rho$, specific internal energy $\epsilon$, $4$-velocity $U^{\mu}$, and
isotropic pressure $P$. The relativistic enthalpy is $h = 1 + \epsilon
+ P / \rho$. The pressure is related to $\rho$ and $\epsilon$ via the
equation of state for an ideal gas, $P = \rho \epsilon ( \Gamma - 1)$. The
magnetic field is described by two sets of variables.  The first is the
constrained transport magnetic field $\mathcal{B}^{i} = [ijk] F_{jk}$,
where $[ijk]$ is the completely anti-symmetric symbol, and $F_{jk}$ are
the spatial components of the electromagnetic field strength tensor.
From the constrained transport (CT) magnetic field components, we
derive the magnetic field four-vector,
$(4\pi)^{1/2}b^{\mu} = *F^{\mu \nu} U_{\nu}$,
and the magnetic field scalar,
$||b^{2}|| = b^{\mu} b_{\mu}$. The electromagnetic component of the
stress-energy tensor is $T^{\mu \nu}_{\mathrm{(EM)}} = \frac{1}{2}g^{\mu
\nu} ||b||^{2} + U^{\mu} U^{\nu} ||b||^{2} - b^{\mu} b^{\nu}$.

In all of the simulations, the initial condition for the matter
is an isolated, hydrostatic torus orbiting near the black hole.
The initial state of the three-dimensional simulations was chosen
so as to be consistent with that used in earlier simulations in this
series \cite[see][]{De-Villiers:2003b}.  The matter's adiabatic index,
$\Gamma$ was $5/3$ and the angular momentum distribution was slightly
sub-Keplerian, with a specific angular momentum at the inner edge of
the torus (located at $r=15$M) $\ell_{\rm in} = 4.567$ (we define $\ell
\equiv U_\phi/U_t$).  For this choice of $\ell_{\rm in}$, the pressure
maximum is at $r \approx 25M$.

In the two-dimensional simulations, our fiducial torus was chosen
to have a constant angular momentum distribution with $\ell = 5.2$,
$\Gamma$ was 4/3, and the inner edge was again located at $r=15$M.
This torus is significantly thicker than the nearly-Keplerian torus
used for the three-dimensional simulations; its density scale height
at the pressure maximum, measured along surfaces of constant radius,
is a factor $\sim1.3$ greater than that of the three-dimensional torus.
This thicker configuration makes it easier to resolve the multiple field
loop model, where several field loops must fit into a restricted space.
This specific combination of $\ell$ and $\Gamma$ was also necessary to
manipulate the shape of the torus so that there was an appropriate balance
in magnetic flux between the different components of the multiple dipole
loop simulation.  In the quadrupolar field case, in order to study how
the evolution of the field depends on its detailed geometry, we examined
what happened as $\ell$ varied from 4.9 to 5.2 in steps of 0.05.

The initial torus is overlaid with a variety of magnetic field configurations,
all with zero net flux.
The initial poloidal magnetic field is determined from the curl of the
four-vector potential, i.e., $F_{\mu\nu} = \partial_{\mu} A_{\nu} - 
\partial_{\nu} A_{\mu}$ with only $A_\phi \neq 0$.  For a single
dipolar loop, we set
\begin{equation}
A_{\phi} = A_0 \left(\rho - \rho_{cut} \right).
\end{equation}
For a quadrupole, the azimuthal component of the vector potential is
\begin{equation}
A_{\phi} = A_0 \left(\rho - \rho_{cut} \right) r \cos \theta.
\end{equation}
In both of these cases $A_\phi$ is set to zero where the density is less
than $\rho_{cut}$; this device confines the initial field to lie well
inside the torus. In all the poloidal simulations, the initial field
intensity (i.e., the parameter $A_0$) is determined by setting
the volume-averaged ratio of gas to magnetic pressure $\beta = 100$.

The vector potential for $n$ dipolar loops (used only for the 
two-dimensional simulations) is more complicated.  It is given by
\begin{equation}
A_{\phi} = A_0 \left(\rho - \rho_{cut} \right) \displaystyle\sum_{\mathrm{j} = 1}^{n}
\frac{J_\mathrm{j} R_\mathrm{j}}{\sqrt{R^{2}_\mathrm{j} + r^{2}
+ 2 R_\mathrm{j} r \sin \theta}} \left[\frac{\left( 2- k_j^{2} \right)
K(k_j) - 2E(k_j)}{k_j^{2}} \right],
\end{equation}
where the total vector potential is the superposition of the
(non-relativistic) vector potentials from $n$ current loops
\cite[see][Pg. 182, Eqn. 5.37]{Jackson:1975}.  Here $J_{\mathrm{j}}$
is the current in the $j$th loop, located at radius $R_{\mathrm{j}}$
in the equatorial plane.  $K$ and $E$ are the complete elliptic
integrals of the first and second kinds, and $k_j = 4 R_\mathrm{j}
r \sin \theta / (R^{2}_\mathrm{j} + r^{2} + 2 R_\mathrm{j} r \sin
\theta )$.  The $R_\mathrm{j}$ are chosen so that the difference
between the inflow times of adjacent loops is approximately equal to
one jet equilibration time, $\sim 700$--$1000M$.  The signs of the $J_j$
alternate with increasing radius so that the senses of the dipolar
field loops alternate.  The magnitudes of the currents were chosen with the goal of
ensuring that the evolution of the two radially innermost loops was governed by the field
local to those loops.  The purpose of this simulation was to study the interaction
of the two inner loops as they accrete; in that context, the only function of the
two outer loops is to supply the outer part of the torus with magnetic field, so
that it could accept the angular momentum brought to it from the true accretion flow.
Thus, the most appropriate outer field structure for this particular numerical experiment
is one that has the least influence on dynamics inside the pressure maximum.
After some experimentation, we found that this goal could be achieved only by
giving the two outer loops relatively small currents.    When that was done,
the outer two loops moved outward with the outer portion of the disk and were quickly
destroyed (i.e. forced to reconnect) as the second loop from the inside expanded
toward them.    We make no pretense that this choice is a fair representation of generic
four-loop evolution, but it does support our study of what happens when two complete
loops of opposite sense are brought into the inner  disk one after the other.

For a purely toroidal initial field, all that is necessary is to set
${\mathcal B}^r = {\mathcal B}^\theta = 0$ and determine ${\mathcal
B}^\phi$ by a choice of volume-averaged $\beta$.  Here we use $\beta=10$,
as, in the purely toroidal case, accretion can occur only once MRI-induced
turbulence on the large scales has grown to sufficiently large amplitude,
and long wavelength modes grow comparatively slowly.  Beginning with a
comparatively strong field ensured that accretion activity started soon
after the simulation began.  After saturation, the field (in relative
terms) actually weakened somewhat: the volume-averaged $\beta$ grew to
$\approx 30$ (see Section \ref{diskbody}). As before, the initial
magnetic field is set to zero for densities below $\rho_{cut}$.

We take the dipole simulation KDPg \cite[originally presented
in][where further discussion of our boundary conditions and grids
may be found]{Hawley:2006} as our fiducial three-dimensional run.
The three-dimensional quadrupole simulation is designated QDPa
and the three-dimensional toroidal simulation is TDPa.  
KDPg and QDPa were run for $10^4\ M$ in time, which corresponds to
approximately $12$ orbits at the initial pressure maximum.  The toroidal
field model, TDPa, evolves initially at a slower rate; it was run
to $2.9\times 10^4 M$ in time.  For each simulation, the time step
$\Delta t$ was determined by the minimal light crossing time for a zone
on the spatial grid and remained constant for the entire simulation
\cite[]{De-Villiers:2003}.

Each of the three-dimensional simulations used $192 \times 192 \times
64$ $(r,\theta,\phi)$ grid zones.  The radial grid extends from an inner
boundary located at $r_{in} = 1.5 M$, which lies just outside the black
hole event horizon, to an outer boundary located at $r_{out} = 120 M$.
For the two poloidal topologies, the radial grid was graded using a
hyperbolic cosine distribution; for the toroidal case a logarithmic
distribution was used.  An outflow condition is applied at both the inner
and outer radial boundary.  The $\theta$-grid spans the range $0.045
\pi \le \theta \le 0.955 \pi$ using an exponential distribution that
concentrates zones near the equator.  A reflecting boundary condition is
used along the conical cutout surrounding the coordinate axis.  
The $\phi$-grid spans the quarter plane, $0 \le \phi \le \pi / 2$, with
periodic boundary conditions applied in $\phi$. The use of this restricted
angular domain significantly reduces the computational requirements of
the simulation.

\begin{figure*}
\leavevmode
\begin{center}
\includegraphics{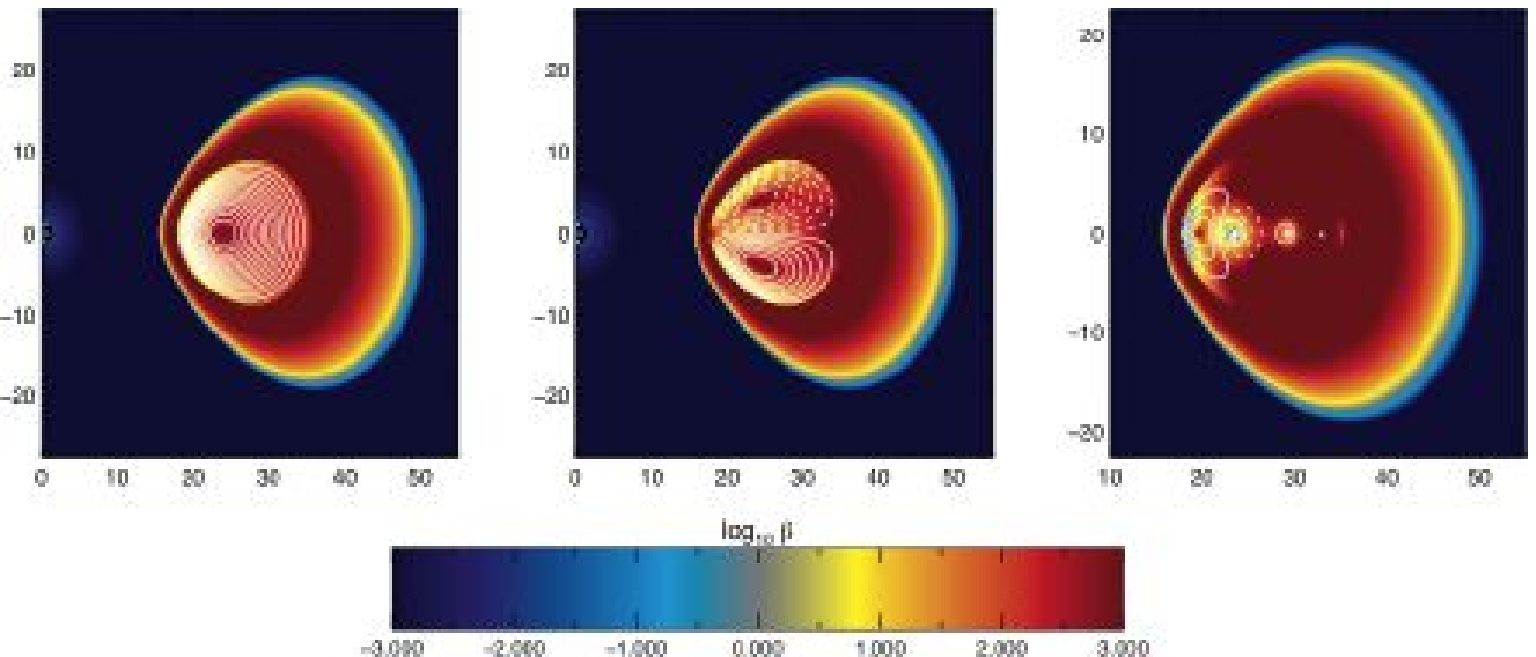}
\end{center}
\caption[]{Initial configurations of dipole (left panel), quadrupole
(center panel) and multiple loop (right panel) field topologies. The torus for
the multiple loop topology is shown slightly zoomed to illustrate better the
initial field structure.  White contours denote magnetic field lines, color contours
the gas $\beta$ parameter.  Solid versus dashed lines indicate field polarity; solid
lines denote current into the page, dashed lines current out of the page.
}
\label{topologies} 
\end{figure*}

The two-dimensional simulations were performed on a $1024\times 1024$
$(r,\theta)$ grid in which the radial cells are graded logarithmically,
and the polar angle cells are concentrated toward the equator in the
same way as for the three-dimensional simulations.  The $\theta$ grid,
however, goes much closer to the axis than in the three-dimensional
simulations, with a cut-out of only $10^{-5}\pi$~radians.  In addition
to allowing higher poloidal resolution at a reasonable computational
cost, two-dimensional simulations have the additional advantage that
field-lines can be easily visualized as level surfaces of the vector
potential component $A_\phi$.  Using this approach, we illustrate the
magnetic field topologies for the three different poloidal configurations
in Figure \ref{topologies}.

\section{The Accretion Flow}\label{flowstructure}

We begin by examining the average properties of the accretion flows in
simulations KDPg, QDPa and TDPa.  \cite{De-Villiers:2003b} described the
accretion flow structure in terms of three distinct regions:
the main disk body, the coronal envelope, and the unbound outflow.
In this section, we study the first two, paying special attention to
contrasts (or lack of contrast) when the magnetic topology changes.
We will discuss the unbound outflow in the following section.

Much of the analysis will be in terms of various time- and space-averaged
quantities.  The radial profile of the time-averaged shell integral of $\cal Q$
is defined as
\begin{equation}
\langle{\cal Q}(r)\rangle_{F} = 
\frac{2}{\pi T} \int \sqrt{-g} {\cal Q}(t,r,\theta,\phi) dt d\theta d\phi
\end{equation}
Similarly, the shell average of a quantity $\cal Q$ at a radius $r$ is defined to be:
\begin{equation}
\langle{\cal Q}(r)\rangle_{S} = 
\frac{ \int \sqrt{-g} {\cal Q} (t,r,\theta,\phi) d\theta d\phi} {\int \sqrt{-g} d\theta d\phi}
\end{equation}
The angular profile of the time-average of $\cal Q$ at radius $r$ 
is defined by
\begin{equation}
\langle{\cal Q}(\theta;r)\rangle_{A} = 
\frac{2}{\pi T}\int \sqrt{-g} {\cal Q} (t,r,\theta,\phi) dt  d\phi.
\end{equation}
Lastly, the time- and volume-average of $\cal Q$ is:
\begin{equation}
\langle{\cal Q}\rangle_{V} = \frac{\int \sqrt{-g} {\cal Q} (t,r,\theta,\phi) dt dr d\theta d\phi}
{\int \sqrt{-g} dt dr d\theta d\phi}.
\end{equation}
In these equations $T$ is the time over which the integral is computed, and
$g$ is the usual metric determinant.  Typically $T=6000M$; for KDPg
and QDPg this is the last $6000M$ of the full evolution, while for
TDPa we choose a $6000M$ window in the middle of the simulation after
the accretion flow is established.  The spatial extent of the shell
integration is the full $\theta$ and $\phi$ computational domain.
During a given simulation, various shell integrals and radial fluxes are
computed and stored every $M$ in time.  This data can then be integrated
over time to obtain quantities such as the total or time-averaged jet
outflow or accretion rate.

It is also useful to divide the shell and volume
integrals into two parts, one for bound and one for unbound flow.
For simplicity we define a particular zone to be ``unbound" if $-h U_t >
1$.  Unbound outflow can further be defined as those unbound cells
with $U^r > 0$.  In these simulations only the outflow near the axis
(the jet outflow) is unbound; the coronal backflow from the disk itself
remains bound.

\subsection{Disk body}\label{diskbody}

The initial evolution of the accretion disks in KDPg and QDPa is
qualitatively similar.   Both field configurations begin with considerable
radial field within the torus.  This is sheared out, generating toroidal
field which, by $t\sim 500M$, is sufficiently strong that the
resulting poloidal gradient in $||b||^2$ begins to drive the inner edge
of the torus (initially located at $r=15M$) inward.  The inner edge of the
disk arrives at the black hole at $t\sim 1000M$.  Within the disk body, the
MRI generates the turbulence that will determine the subsequent evolution
of the disk, and by $t\sim 4000M$ a statistically stationary
turbulent accretion flow has been established inside the radius of
the inner edge of the initial torus.

The toroidal field model TDPa evolves more slowly than the two poloidal
field cases, consistent with the results from earlier toroidal field
pseudo-Newtonian simulations \citep{HK:2002}.  As discussed at length in
\cite{HK:2002}, this behavior stems both from the absence of an initial
radial field (which means that there is no toroidal field amplification
due to shear) and from the fact that long-wavelength modes---which
are the most effective in driving accretion---grow relatively slowly.
Inflow can begin only when the MRI has produced turbulence of sufficient
amplitude, which occurs by $t=4000M$, corresponding to about 5 orbits at
the radius of the torus pressure maximum.  The accretion rate into the
hole increases until about $t=1.5\times 10^4 M$, after which it shows
large fluctuations without an overall trend.

\begin{figure*}
\begin{center}
\leavevmode
\includegraphics[width=0.44\textwidth]{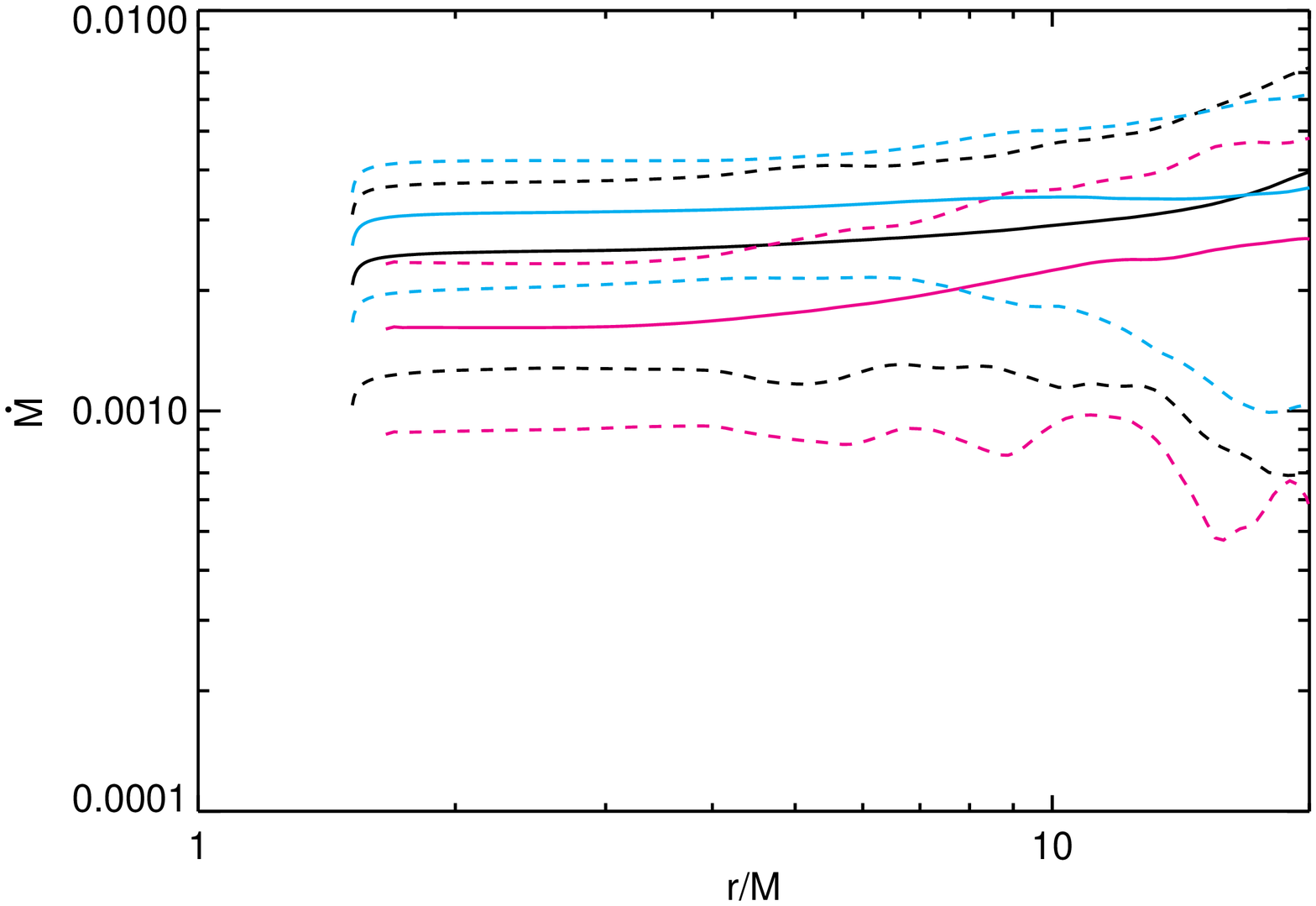}
\includegraphics[width=0.44\textwidth]{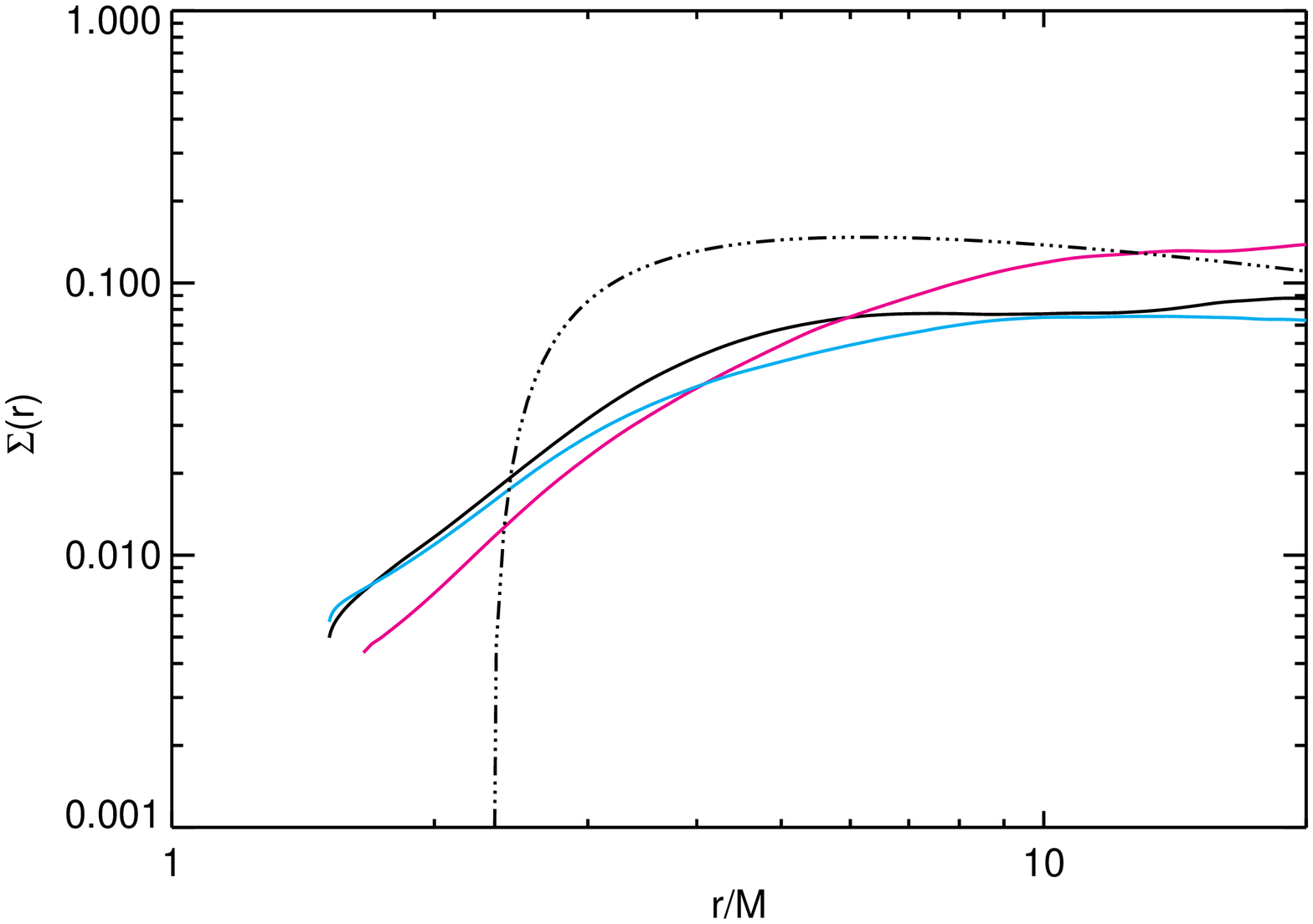}
\includegraphics[width=0.44\textwidth]{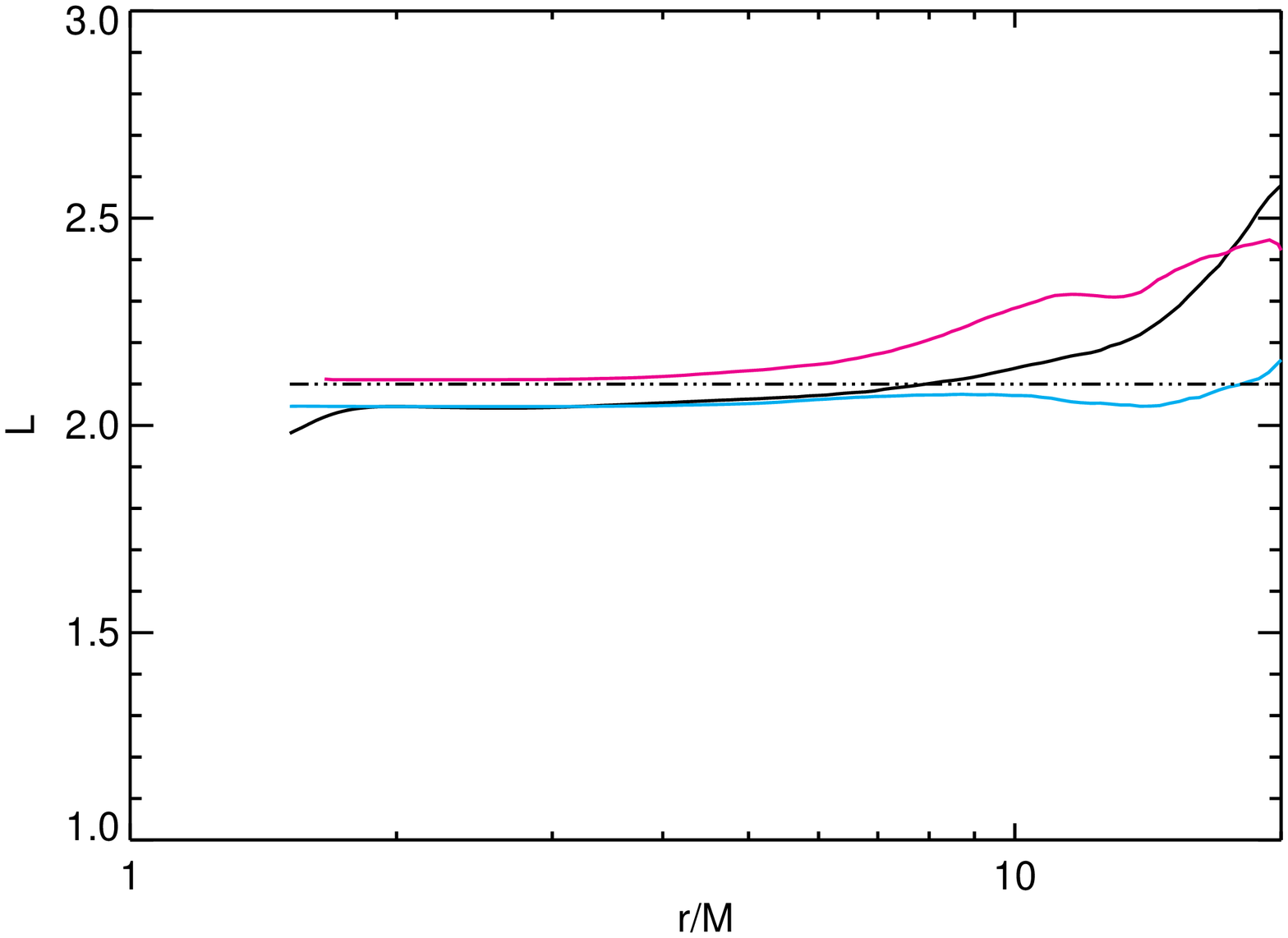}
\includegraphics[width=0.44\textwidth]{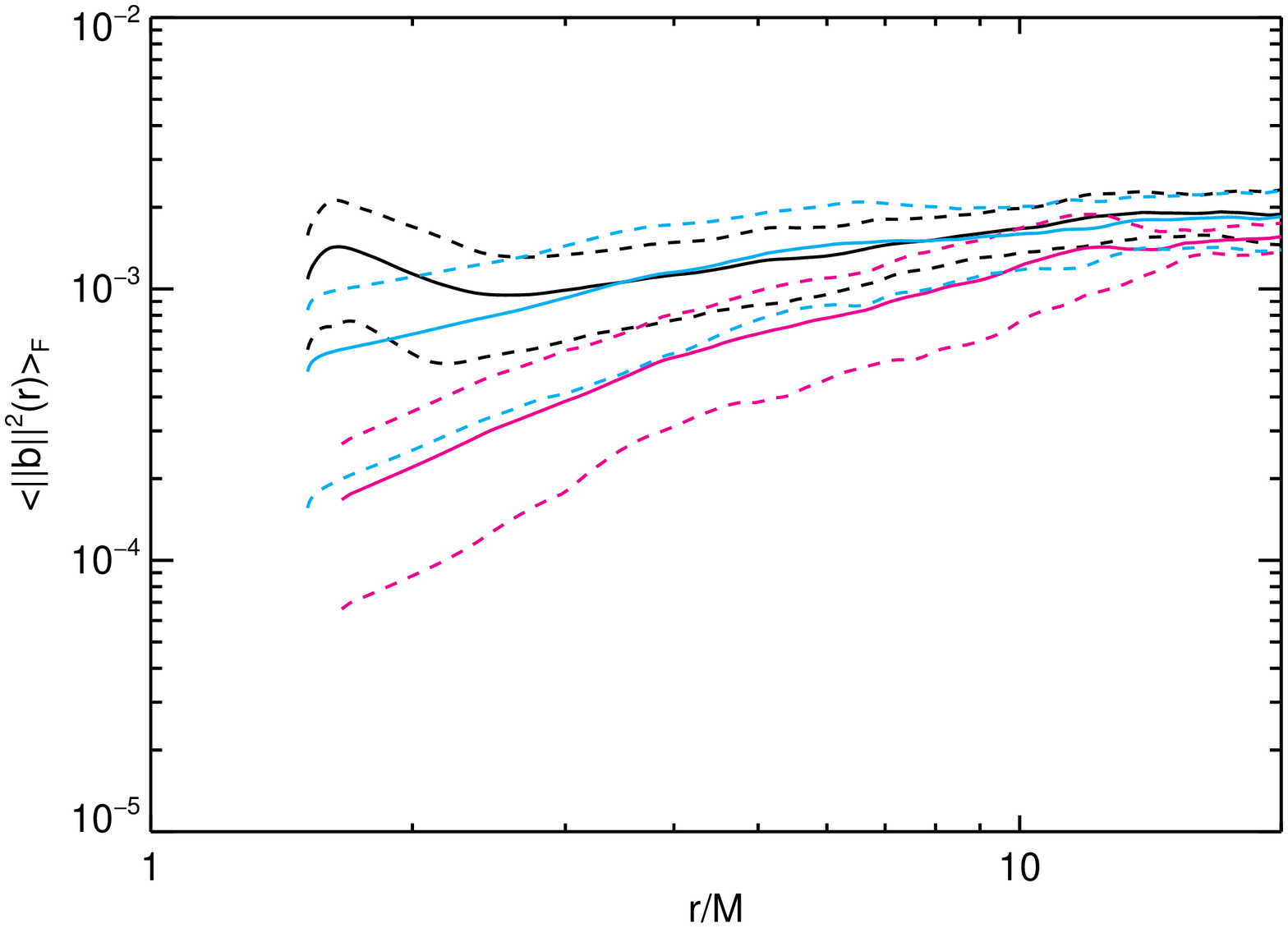}
\includegraphics[width=0.44\textwidth]{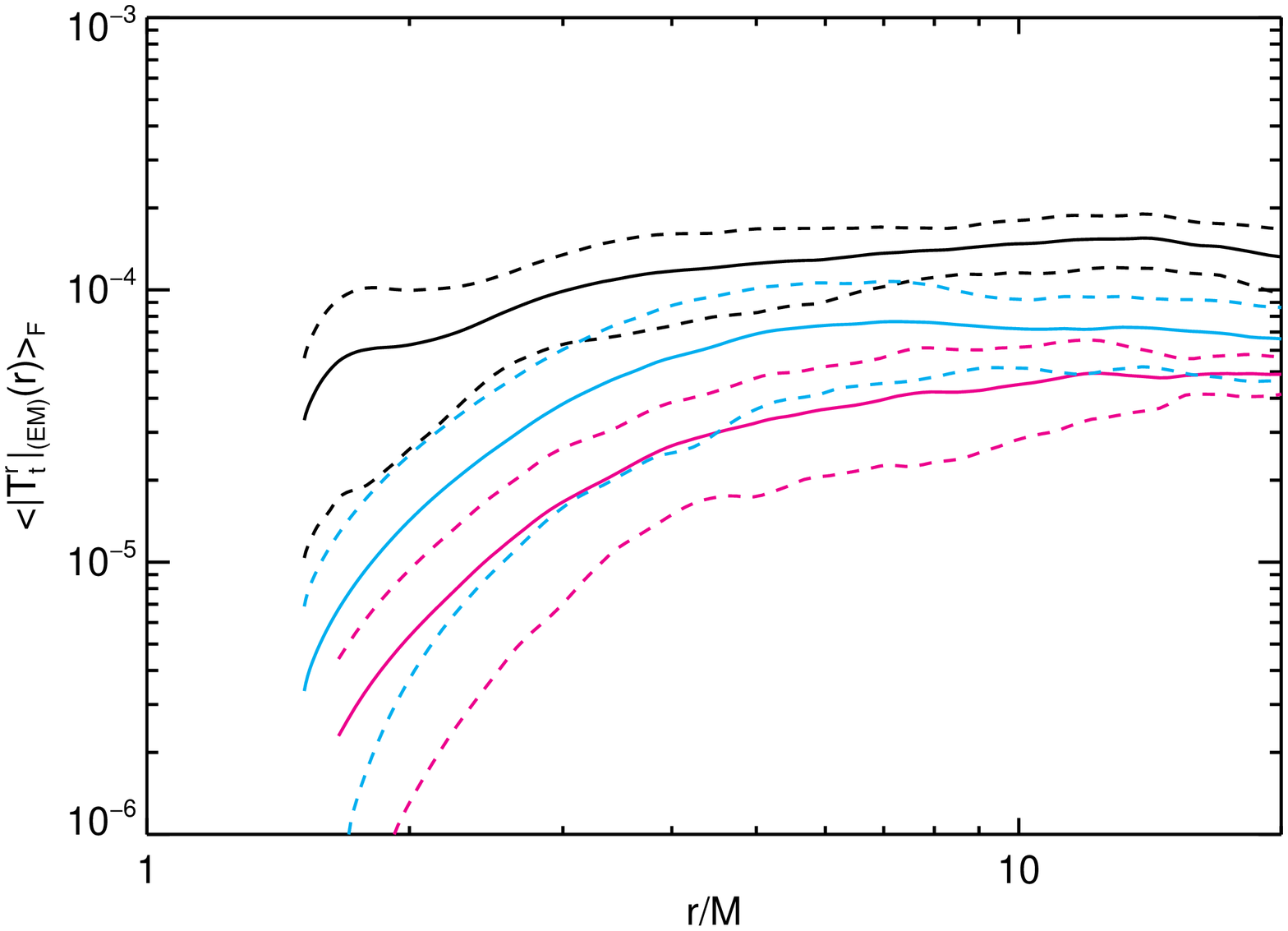}
\includegraphics[width=0.44\textwidth]{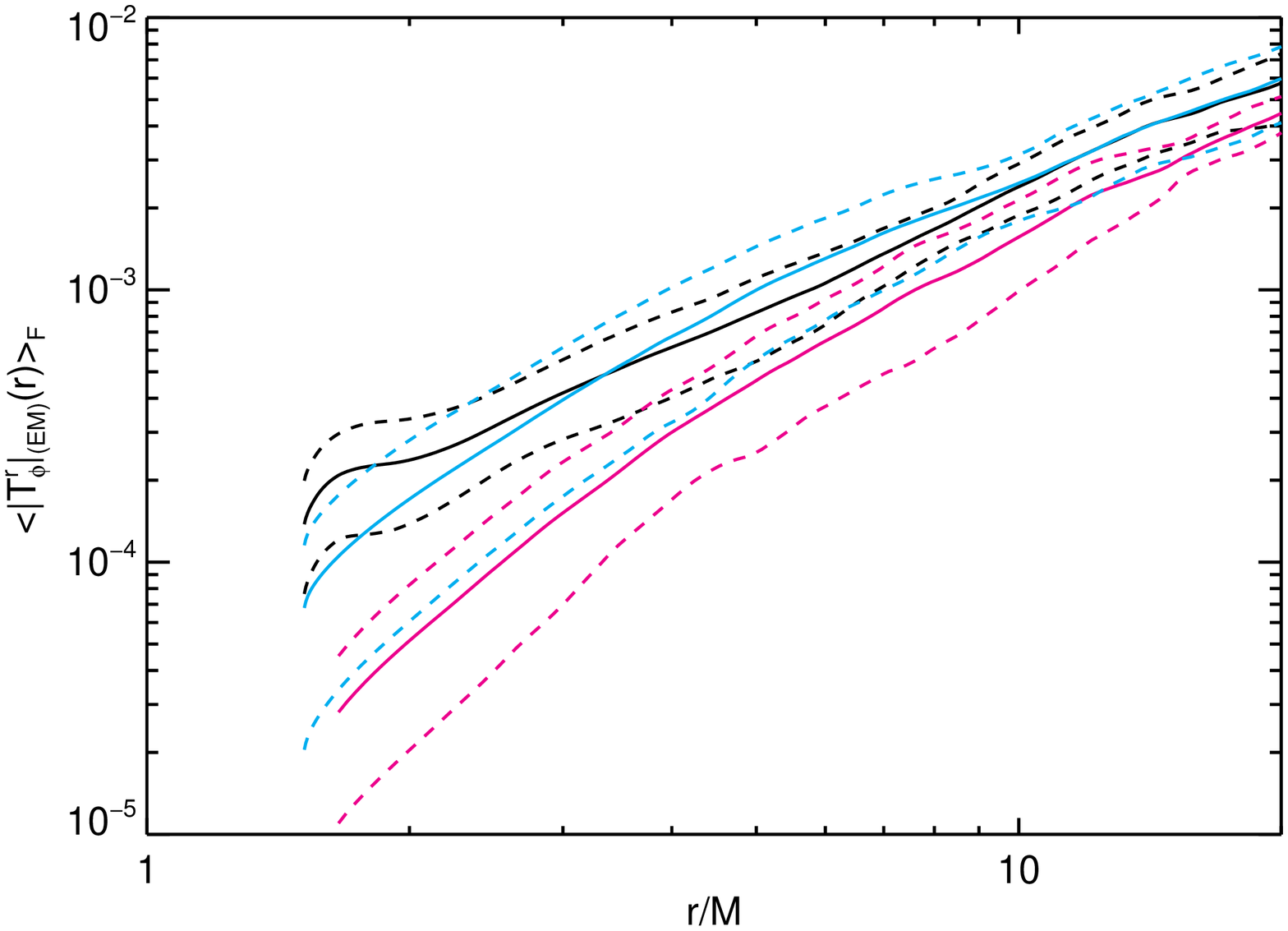}
\end{center}
\caption[]{Time-averaged, shell-integrals of the accretion rate, $\dot{M}$ (top left panel), surface density $\Sigma(r)$ \cite[top right panel, dot-dash lines show the surface density distribution predicted by][]{Novikov:1973}, the net accreted angular momentum
per unit rest mass, $L$, (center left panel; dot-dashed line indicates value at the ISCO), magnetic energy density $\langle ||b||^2 (r) \rangle_{F}$ (center right panel); the electromagnetic contribution to the energy flux $\langle |T^r_{t} |_{\; \mathrm{(EM)}} (r) \rangle_{F}$ (bottom left panel) and angular
momentum flux $\langle |T^r_{\phi} |_{\; \mathrm{(EM)}}  (r) \rangle_{F}$ (bottom right panel) in bound material.  Each figure shows data from KDPg (black lines), QDPa
(blue lines) and TDPa (magenta lines) time-averaged over $4000-10000$M
for KDPg and QDPa, and $12500-18500$M for TDPa. Solid lines denote the
time average, dashed lines (where shown) $\pm1$ standard deviation from the
average. Note that the radial coordinate of the ISCO is $2.32M$ for the black
hole spin $a/M = 0.9$.}
\label{bndavgplt1} 
\end{figure*}

Figure~\ref{bndavgplt1} shows time-averaged, shell-integrated radial profiles of a
number of quantities relevant to the accretion flow: accretion rate
$\dot{M} = \langle \rho U^{r} (r) \rangle_{F}$, surface density $\Sigma(r) = \langle \rho \rangle_F / \int \sqrt{g_{rr}g_{\phi\phi}} d\phi$, the net accreted angular momentum per unit rest mass, $L = \langle T^{r}_{\phi \; \mathrm{(FL)}} (r) + T^{r}_{\phi \;\mathrm{(EM)}} (r)\rangle_{F} / \dot{M}$, $\langle ||b||^2 (r) \rangle_{F}$, the electromagnetic Poynting flux $\langle |T^r_{t} |_{\; \mathrm{(EM)}} (r) \rangle_{F}$ and the EM angular momentum flux $\langle |T^r_{\phi} |_{\; \mathrm{(EM)}}  (r) \rangle_{F}$.  The subscripts FL and EM denote the fluid and electromagnetic contributions to the stress-energy tensor, respectively.
All quantities are computed in the coordinate frame after the turbulent
accretion flow is established, and in all cases the volume integral
was restricted to cells where the matter was bound.  The poloidal field
simulations were averaged over time $t=4000$--$10000M$ while the toroidal
field data were averaged over $t=12500$--$18500M$.

These six plots may be divided into two groups: those with little
dependence on initial field topology (surface density and accreted angular
momentum per unit mass), and those with a stronger dependence (accretion
rate, magnetic field strength, Poynting flux, and electromagnetic angular
momentum flux).  Although the radial distribution of the surface density
in all three cases is fairly similar, there is a slight distinction
between the two poloidal models on the one hand and the toroidal on
the other.  The surface density in the toroidal model rises somewhat
more steeply with radius, with lower density near the black hole and
higher values outside of $r/M=7$.  For all three topologies, the specific
angular momentum carried into the hole is close to the angular momentum
of the marginally stable orbit.  For TDPa, the value is essentially
equal to that of the ISCO, while both QDPa and KDPa have smaller values,
indicative of continued stress operating near or inside the ISCO.

In all three cases, the time-averaged accretion rate is constant with
radius out to $r/M \simeq 7$--10; in this sense, all three have achieved a
quasi-steady state within $\simeq 4 r_{\rm ISCO}$, where $r_{ISCO}$ is the
radial coordinate of the innermost stable circular orbit, $2.32M$
(the rise in $\dot{M}$ and $L$ at large radius in TDPa is likely an
artifact of the stronger magnetic field with which that simulation
was begun; equilibration takes place on timescales proportional to
the local orbital period, which is, of course, $\propto r^{3/2}$, so
inflow equilibration is slower at large radius).
However, the actual value of the accretion rate is significantly
different in the three simulations.
The time-averaged mass accretion rate is highest in QDPa, exceeding that
of KDPg by about $30\%$, while $\dot M$ is lower in TDPa by more than
a factor of 2 compared to QDPa. 

Accretion is, of course, driven by angular momentum transport, which
is largely due to magnetic torques.  To support these torques, the
shell-integrated electromagnetic angular momentum flux must increase
outward, as the net torque
on the matter is the divergence of the EM angular momentum flux. The
dependence of the electromagnetic angular momentum flux on initial
field topology is therefore similar to that of the mass accretion rate:
it is consistently greater for KDPg and QDPa than for TDPa.  The most
significant difference between the KDPg and QDPa curves is that in
the region near the ISCO, the electromagnetic angular momentum flux
in KDPg becomes almost constant with decreasing radius; this effect
has previously been seen in other simulations that begin with dipole
magnetic field loops \citep{Krolik:2005}.  In the other two cases,
the EM angular momentum flux continues to decline inward.

The mean magnetic field strength and Poynting flux show a very similar
dependence on initial field topology: QDPa and KDPg closely resemble
each other, TDPa is somewhat weaker.  In the toroidal case, the disk body
average magnetic energy density is weaker by a factor of several than in
the other two simulations, and decreases inward toward the horizon rather
more rapidly.  Correspondingly, the time- and volume-averaged $\beta$
parameter was 11 in QDPa, 16 in KDPg, and 30 in TDPa (because this is
a volume-average, the corona is weighted heavily in this statistic).
The electromagnetic energy flux shows the greatest variation between
models.   $\langle |T^r_{t} |_{\; \mathrm{(EM)}}  (r) \rangle_{F}$ in KDPg is consistently larger than the value in QDPa: by a factor of two at $r/M=10$ and a factor of
six at $r/M=1.65$.   The value in TDPa is smaller still by a factor of
$\sim 2-4$ at these same radii.

\begin{figure*}
\begin{center}
\leavevmode
\includegraphics[width=0.32\textwidth]{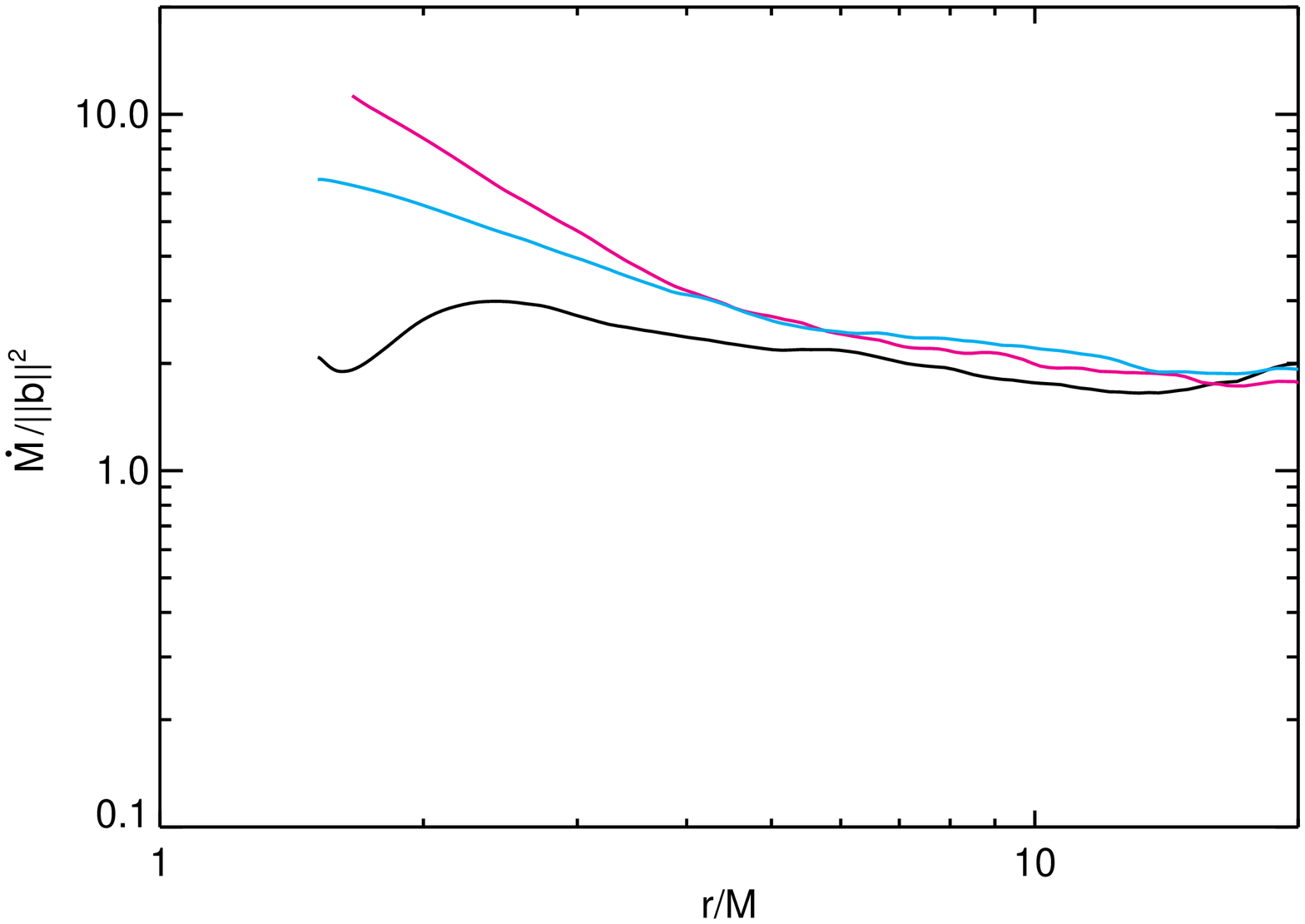}
\includegraphics[width=0.32\textwidth]{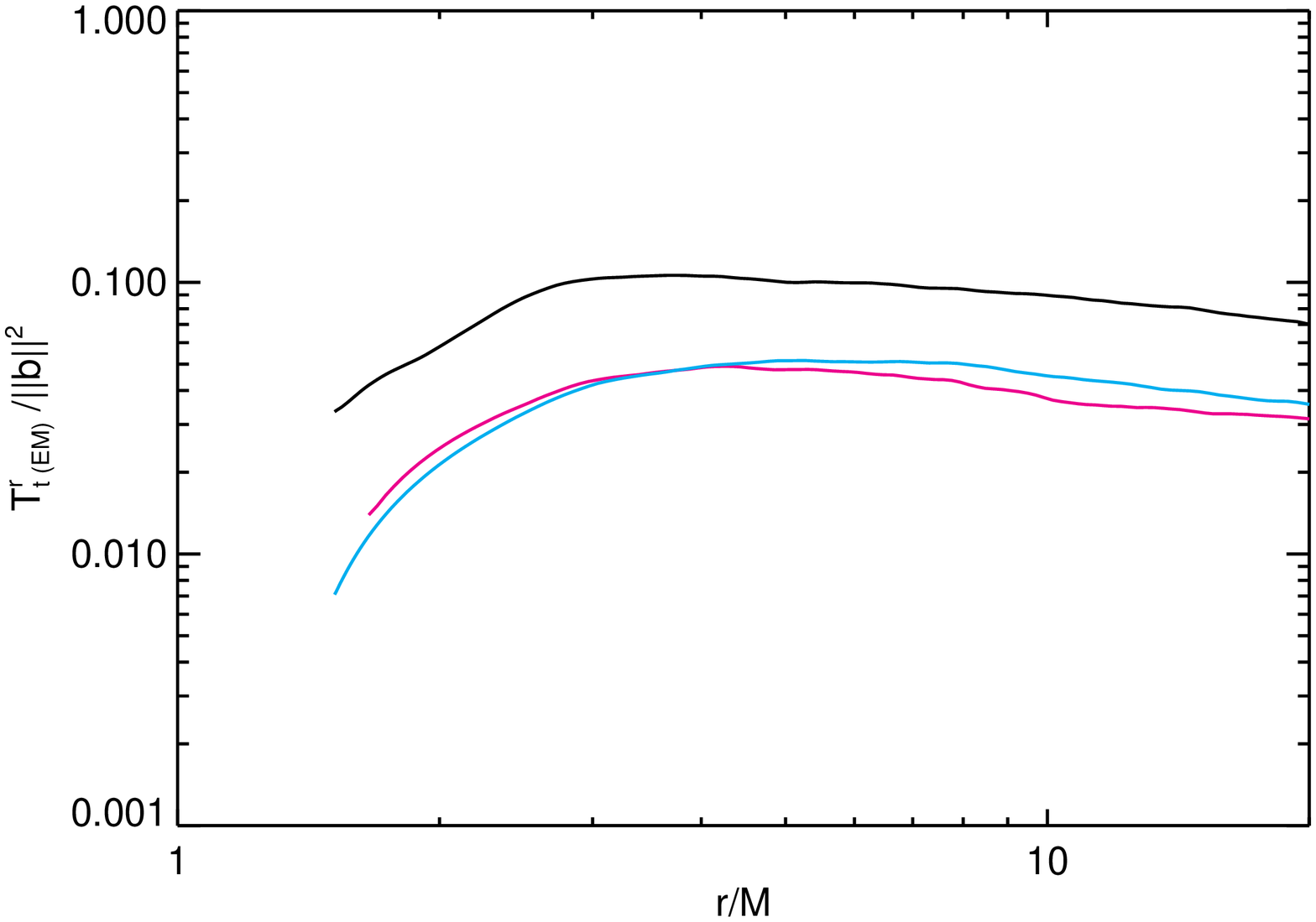}
\includegraphics[width=0.32\textwidth]{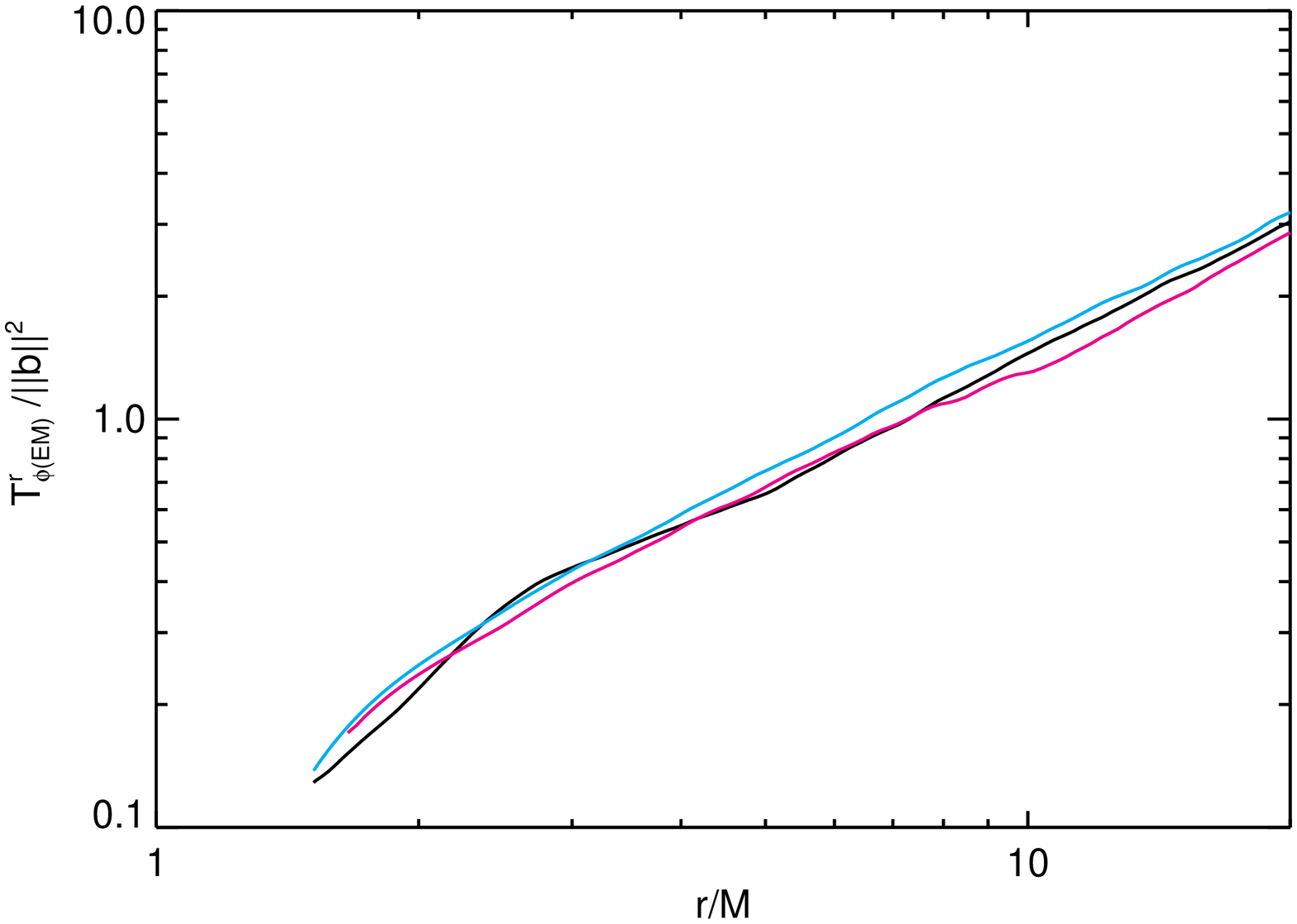}
\end{center}
\caption[]{Time-averaged ratio of the accretion rate $\dot{M}$ (left
panel), the electromagnetic contribution to the energy
flux $\langle |T^r_{t} |_{\; \mathrm{(EM)}} (r) \rangle_{F}$ (center panel) and angular
momentum flux $\langle |T^r_{\phi} |_{\; \mathrm{(EM)}}  (r) \rangle_{F}$ (right panel) to the magnetic field strength $\langle ||b||^2 (r) \rangle_{F}$; all are restricted to bound material.  Each figure shows data from KDPg (black lines), QDPa
(blue lines) and TDPa (magenta lines).  The time-averaging is over the
period $4000-10000$M for KDPg and QDPa, and $12500-18500$M for TDPa. }
\label{ratio} 
\end{figure*}

All these relations can be understood in a straightforward fashion:
there is a nearly-constant ratio between the EM angular momentum
flux (and therefore the accretion rate), the Poynting flux, and the
magnetic field intensity.  In Figure~\ref{ratio}, we plot the ratios of
$\dot{M}$, $\langle |T^r_{t} |_{\; \mathrm{(EM)}}  (r) \rangle_{F}$ and$\langle |T^r_{\phi} |_{\; \mathrm{(EM)}}  (r) \rangle_{F}$
to $\langle ||b||^2 (r) \rangle_{F}$.  The inter-simulation contrast in the means of these
quantities falls by a factor of several from the plot of their absolute
values (Fig.~\ref{bndavgplt1}) to the plot of their ratios to $||b||^2$
(Fig.~\ref{ratio}).   Thus, we see that their primary dependence is on
the magnetic field strength.  Only in the ratio of the electromagnetic
energy flux to magnetic field strength do we see any global variation
between the field topologies: $\langle |T^r_{t} |_{\; \mathrm{(EM)}}  (r) \rangle_{F}$ in KDPg is consistently a factor of two above QDPa and TDPa.  This contrast arises
from the strong Poynting flux in KDPg associated with the slower-moving,
bound portion of the funnel-wall outflow \cite[as can be seen from Figure
10 of][see \S\ref{jets}]{Krolik:2005}.

\begin{figure*}
\begin{center}
\leavevmode
\includegraphics[width=0.32\textwidth]{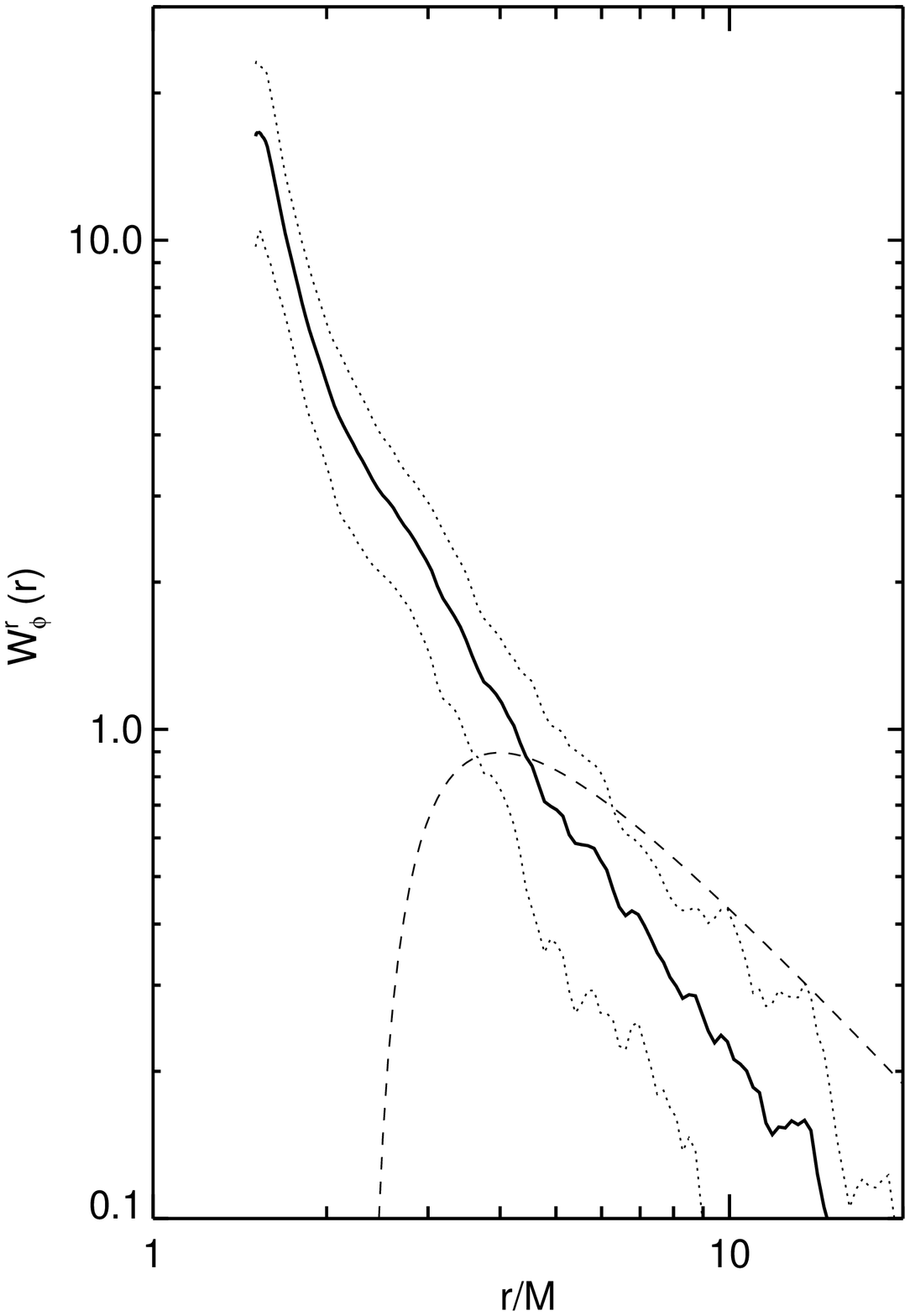}
\includegraphics[width=0.32\textwidth]{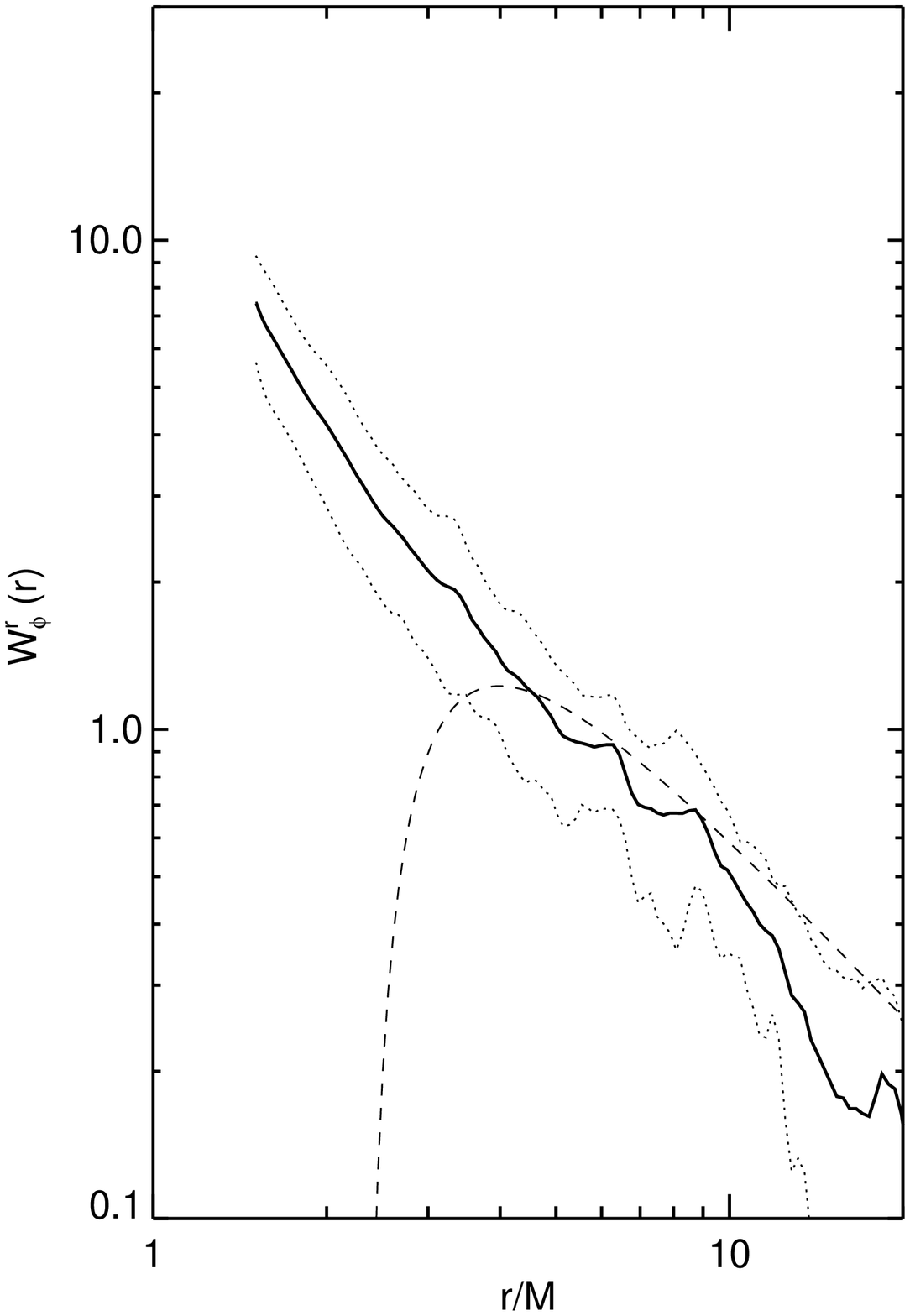}
\includegraphics[width=0.32\textwidth]{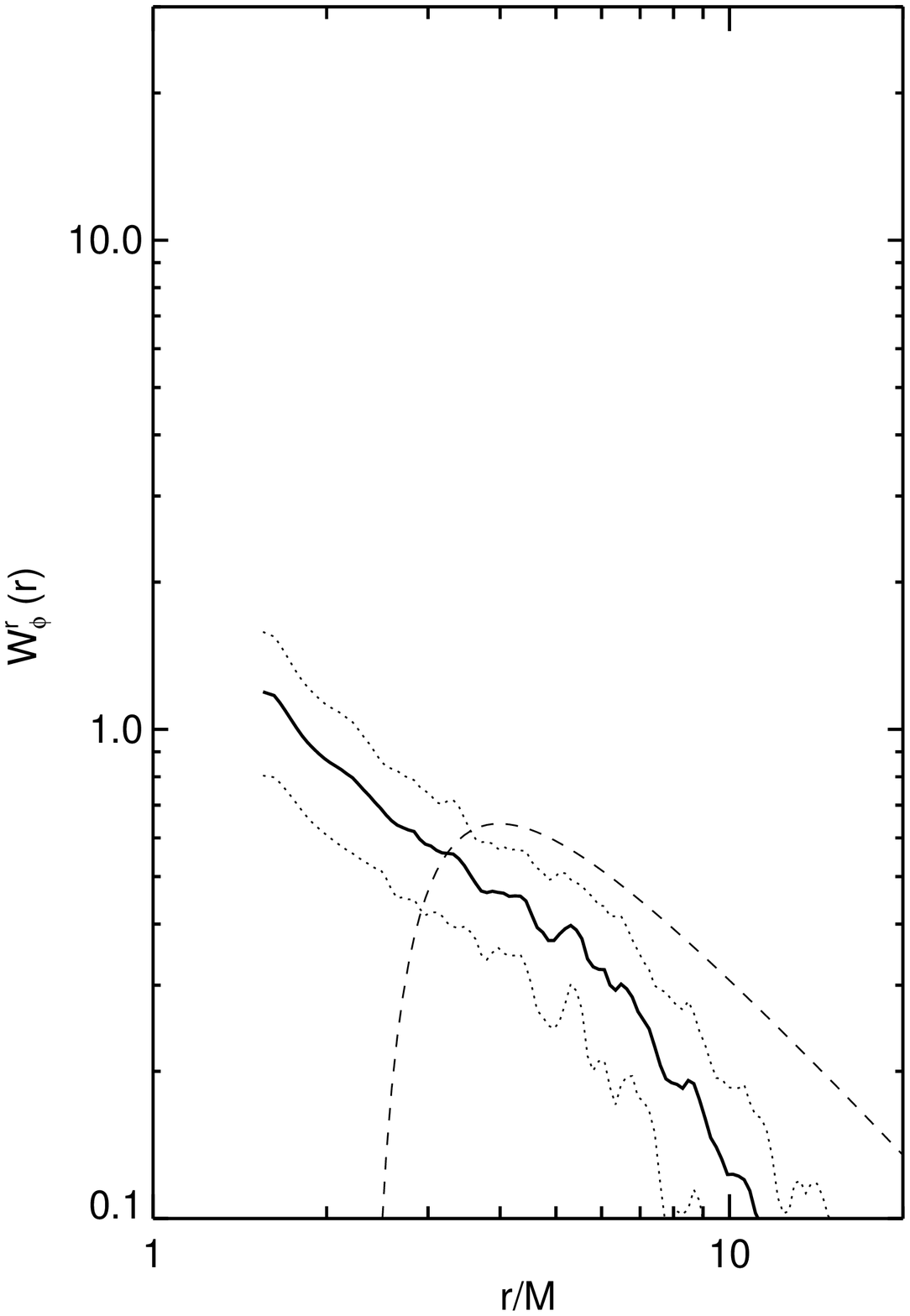}
\end{center}
\caption[]{Time-averaged fluid-frame Maxwell stress, ${\cal W}^{(r)}_{(\phi)}$
(solid lines) for simulations KDPg (left panel), QDPa (center panel) and
TDPa (right panel). Solid lines denote the time average, dotted lines
$\pm1$ standard deviation from the average. The Novikov-Thorne prediction
of the fluid-frame stress for a thin disk with an accretion rate
equal to the time-average for that simulation is shown with a dashed line. KDPg and QDPa are averaged over 8000--$10000 M$ and TDPa is
averaged over 2--$2.2\times 10^4 M$.}
\label{maxwell} 
\end{figure*}

The behavior of the fluid-frame stress can be seen in Figure~\ref{maxwell},
which depicts its time-averaged vertically-integrated value as a function of
radius.  To compute this quantity, we project the stress tensor onto a local
tetrad system in each cell and multiply by the fluid-frame cell volume,
which is computed by a similar projection \citep{Krolik:2005}.  After
integrating over the bound matter on that spherical shell, we normalize
the result to the surface area that this finite-thickness shell occupies
in the equatorial plane.  We present the result in this form in order to
compare with the conventional representation of the vertically-integrated
stress given by \cite{Novikov:1973}, which is presented in these terms.
The Novikov-Thorne fluid-frame stress, unlike ours, is {\it assumed}
to be zero at and within the ISCO.   In the figure, the Novikov-Thorne
stress is normalized by the time-averaged value of the accretion rate in
the simulation.

Outside the ISCO, the radial profile of the Maxwell stress is similar
to the prediction of the Novikov-Thorne model, but in no case is there a
roll-over in the value near the ISCO.  As has been noted previously for
the initial dipole field topology \citep{Krolik:2005}, the $r-\phi$ stress
does not go to zero at the ISCO, but is continuous down to the event
horizon.  This is true for all three field topologies, at least in a
time-averaged sense, although the amplitude of the stress declines as
one goes from dipole to quadrupole to toroidal field topology.  In the
dipole case, the logarithmic gradient of the stress as a function of
radius actually steepens as the accretion flow approaches the horizon.

As can be seen from Figure~\ref{bndavgplt1}, this extra stress does
reduce the specific angular momentum carried into the hole by the
accretion flow.  At the ISCO the accretion flow in the toroidal field
simulation has $L \sim L_{ISCO}$ (where $L_{ISCO} = 2.1$ is the specific
angular momentum of the ISCO) and the accretion flows
in the dipole and quadrupole simulations have $L \sim 2.05$ (a change of
$\sim 2.5\%$ compared to $L_{ISCO}$). Below the ISCO, $L$ in the quadrupole
and toroidal cases remains constant with decreasing $r$.  In the dipole
simulation, $L$ decreases to $\sim 1.98$ at $r/M = r_{in}$, a decrease of
$\sim6\%$ compared to $L_{ISCO}$. The fall off in the dipole case below
the ISCO is presumably due to the steepening of the Maxwell stress in
this region.  

As has been remarked in earlier papers \cite{Krolik:2005},
\cite{Gammie:2004}, when $a/M = 0.9$, the magnitude of the Maxwell stress
(i.e., the electromagnetic angular momentum flux) is smaller than the
angular momentum flux associated directly with matter ($\dot M u_\phi$)
in the inner disk.  It follows (as shown in Figure~\ref{ratio}), that the
mean accreted angular momentum per unit rest-mass is only slightly below
the Novikov-Thorne value (i.e., $u_\phi$ at the ISCO).  Nonetheless,
this continuity in the stress is important for other reasons: because
the dissipation is rather more centrally-concentrated than the stress,
this additional stress can be significant in the disk heat budget; and
the relative importance of the additional stress increases sharply for
higher black hole spins.

 Thus, we see that the initial magnetic field geometry makes relatively
little qualitative difference to the character of the accretion flow.
The sense of its (weak) influence is that the simulation beginning with
a purely toroidal magnetic field results in a saturated field amplitude
a factor of two smaller than those resulting from poloidal seed fields.
The Poynting flux, electromagnetic angular momentum flux, and accretion
rate scale in direct proportion to the magnetic field energy density.

In their study based on axisymmetric simulations, \cite{McKinney:2007} also found
that certain disk properties are largely independent of the initial field configuration.  They
found that the shell-averages of magnetic field quantities within the bound portion of
the accretion flow (strictly speaking, where $||b||^2 < \rho$) follow
power-law distributions that are the same regardless of the initial field  topology.
For example, they found that the field amplitude $\langle ||b||(r) \rangle_{S} \propto r^{-1.3}$  from the ISCO
out to about $r=10M$.  Integrating the same magnetic field quantities (magnetic field strength
and the lab-frame projections of the CT-fields) over the bound portion of the flow
in our simulations reveals that these quantities do scale roughly as power-laws in
radius, but with shallower gradients than as determined by \cite{McKinney:2007}.   In
addition, we see a slow trend with field topology, as well as significant fluctuations in
time:  $\langle ||b||(r) \rangle_{S} \propto  r^{-0.75\pm0.09}$ in simulation KDPg (where $\pm$ denotes 1 std.
deviation about the time-average), $\langle ||b||(r) \rangle_{S} \propto r^{-0.67\pm0.08}$ in  simulation QDPa,
and $\langle ||b||(r) \rangle_{S} \propto r^{-0.58\pm0.09}$ in simulation TDPa.  The contrast between our results and \cite{McKinney:2007} could result from a number of considerations: the initial torus configurations differed in their radial pressure and angular momentum profiles, and our simulations are
fully three-dimensional, rather than axisymmetric.    These scalings may or may not be
universal, but they appear to have little to do with the jet.  As we will shortly show in
\S\ref{jets}, jet properties differ strongly from one initial magnetic topology to another; it
must be, then, that jets depend on other properties in addition to the current distribution in the disk.

\subsection{Corona}

\cite{De-Villiers:2003b} noted that the main disk body is surrounded by
a low density region where the magnetic and thermal energies are in
approximate equipartition.  This region was designated as the coronal
envelope.  In contrast to the main disk body, where poloidal fluid
velocities are dominated by turbulence, the poloidal motions in the
coronal envelope are smooth on small scales and create an outgoing,
but bound, backflow.

\begin{figure*}
\begin{center}
\leavevmode
\includegraphics{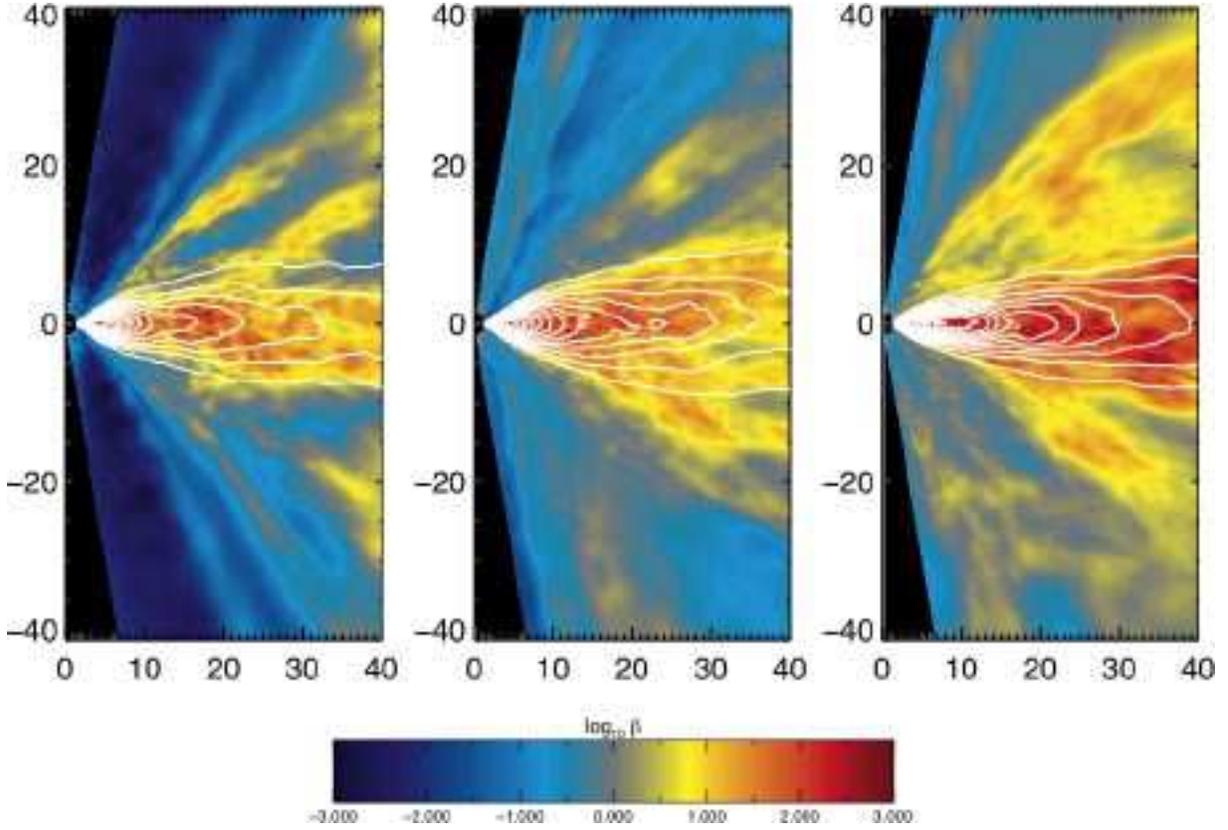}
\end{center}
\caption[]{Color contours of the time- and azimuthally-averaged
$\beta$ parameter for simulations KDPg (left panel), QDPa (center panel)
and TDPa (right panel), overlaid with white contours of gas
density.  KDPg and QDPa are averaged over 8000--$10000 M$ and TDPa is
averaged over 2--$2.2\times 10^4 M$.}
\label{beta} 
\end{figure*}

Figure \ref{beta} displays the spatial distribution of the time- and
azimuthally-averaged $\beta$ parameter overlaid with contours of the
gas density for each of the three different topologies.  KDPg shows the
familiar strongly-magnetized axial funnel and a corona with relatively
large regions of $\beta < 1$.  QDPa shows a lower level of magnetization
in the funnel region, while TDPa has much larger regions with $\beta >
1$ overall.  To quantify these impressions, we adopt a more precise
definition of the corona than that given in \cite{De-Villiers:2003b}.
We define it as those regions of {\it bound} material that lie outside
of $\sqrt{2}$ density-scale heights from the mid-plane, that is those
cells that satisfy both $-h U_t \le 1$ and $\rho / \rho_{eq} \le e^{-2}$,
where $\rho_{eq}$ is the equatorial density and we 
measure along curves of constant $r$.

Using this definition, we integrate over the corona and over time
to compute the average total volume and mass fraction in the corona,
the volume-averaged $\beta$ parameter, and the mean fluid-frame
thermal and magnetic energy per unit mass.  These data are given in
Table~\ref{corona}.

The coronal regions for all three simulations take up half or more of the
total simulation volume but contain only $5\%$ of the mass.  The volume
assigned to the corona for KDPg is a smaller percentage compared to the
other models, but KDPg also contains an unbound jet along the funnel
axis, which is excluded from the corona by definition.
Within the coronal volume, the average thermal and magnetic
energies per mass are similar, with volume averaged $\beta$ values
of $\simeq 2$--4.  

Figure \ref{angcor} plots the angular profiles of density $\langle \rho (\theta; r) \rangle_{A}$, gas pressure $\langle P (\theta; r) \rangle_{A}$, 
magnetic pressure $\langle \frac{1}{2} ||b||^{2} (\theta; r) \rangle_{A}$ and $\beta$ $\langle 2 P / ||b||^{2} (\theta; r) \rangle_{A}$ at $r=10M$.  All three magnetic topology cases
have very similar density profiles in the disk body, but appear to differ
in the corona.  Although our averaging period of $6000M$ is 30
orbital periods at the radius shown ($r = 10M$), it is not long
enough to erase completely fluctuations in the coronal density profile:
observe that the quadrupolar case is the most extended in the ``southern
hemisphere" corona while the toroidal case is the most extended in
the ``northern hemisphere".

Consistent with the general behavior we have already emphasized,
the gas and magnetic pressure profiles in the disk body are very
similar in the dipolar and quadrupolar cases, but both pressures are
rather smaller in the toroidal field case.  Nonetheless, all three cases
share one important property: the magnetic pressure is almost flat
for several scale-heights around the midplane.   This finding is
consistent with results from shearing-box simulations with much
better vertical resolution \citep{Hirose:2005,Hirose:2007}.
In the corona, the magnetic pressure drops rapidly with height in
QDPa and TDPa, but remains nearly constant in KDPg.
In all three cases, the magnetic and gas pressures track each other closely,
so that the $\beta$ profiles in the three runs are very similar.

Figure \ref{angtrptrt} plots the angular profiles of the electromagnetic
contributions to  $T^r_\phi$ and $T^r_t$ ($\langle |T^r_{t} |_{\; \mathrm{(EM)}} (\theta; r) \rangle_{A}$ and $\langle |T^r_{\phi} |_{\; \mathrm{(EM)}} (\theta; r) \rangle_{A}$ respectively) at two different radii,
$r/M=1.65$ and $r/M=10$.  At the larger radius, it should come
as no surprise that the profile of magnetic pressure shown at that
radius in Figure~\ref{angcor} accurately predicts the profile of these
two electromagnetic quantities.  At the smaller radius, as we have
already shown (Fig.~\ref{bndavgplt1}), the magnetic field is considerably
more intense in the dipole case than in the other two, so both
the electromagnetic angular momentum flux and energy flux follow
suit.

To summarize this section, we have found that, just as for the main
disk body, the character of disk coronae is only weakly dependent upon
the initial field topology.  There are modest quantitative contrasts
of the same sort as seen in the disk body---dipolar and quadrupolar
initial fields both create somewhat stronger magnetic fields on average
than does initial toroidal field---but that is all.

\section{Outflows}\label{jets}

\subsection{Global measures}

The formation of unbound outflows, or jets, is one of the most striking
features of the KD models \cite[]{De-Villiers:2003b}, all of which were
initialized with single-loop dipolar magnetic fields.  These outflows
generically have two components:  a Poynting flux jet inside the axial
funnel formed by the centrifugal barrier, and a sheath of much denser
unbound gas moving outward at sub-relativistic speed  along the funnel's
outer edge.  As described in \cite{Hawley:2006}, energy for the Poynting
flux comes from the rotation of the black hole, in a manner closely
related to the classical Blandford-Znajek mechanism.  In the case of
the Schwarzschild hole, the absence of black hole rotation means that
the magnetic field in the funnel, although relatively strong, is mostly
radial and carries no Poynting flux.  The funnel wall jet, on the other
hand, appears to be powered by the significant gas and magnetic pressure
near the ISCO.  The gas retains enough angular momentum to be excluded
from the funnel and is squeezed outward along the centrifugal barrier.

In contrast, the three-dimensional quadrupole and toroidal
field simulations  have substantially weaker unbound outflows, if any.
Figure \ref{ubndavgplt1} shows the time-averaged shell-integrals of the
mass outflow rate $\langle\dot{M}(r)\rangle_{F}$, magnetic field strength$\langle ||b||^2 (r) \rangle_{F}$; along with the electromagnetic energy flux, $\langle |T^r_{t} |_{\; \mathrm{(EM)}} (r) \rangle_{F}$ and angular momentum flux $\langle |T^r_{\phi} |_{\; \mathrm{(EM)}} (r) \rangle_{F}$ in unbound
material between $r=10$M and $r=100$M. The values in KDPg are 10--100
times larger than those in QDPa, whose outflow quantities are, in turn,
an order of magnitude larger than those in TDPa.  For example, the unbound
mass outflow rate at $r=100M$ in KDPg is $\simeq 25$ times that of QDPa,
while the mass outflow rate in TDPa is 10 times smaller than in QDPa.
The strength of the outflow is directly related to the strength of the
funnel field.  In KDPg the time-averaged funnel magnetic energy density
is 20--$400$
times greater than in the quadrupole model, while the toroidal model is
weaker by another factor of 10.  The mean Poynting flux in KDPg is therefore
$\sim 100$ times the Poynting flux for QDPa.  In other words, unlike
the qualitative similarity of the accretion flows produced by these
different field topologies, the outflows differ dramatically.  To
understand better the origin of this sharp contrast, in the next section
we focus more closely on exactly how field is introduced into the funnel,
and how its magnitude is controlled.

\subsection{Funnel Field Creation}\label{earlytimes}

We have seen that the strength of the Poynting flux jet is very sensitive
to the initial magnetic field topology.  Because the establishment of
large-scale field within the funnel appears to be critical to driving
a jet with substantial Poynting flux, we devote the remainder of this
section to a close examination of just how different magnetic topologies
lead to different funnel magnetic fields, leaning heavily on high-resolution
2-d axisymmetric simulations.

\subsubsection{Dipole Topology}\label{dipole}

We begin by analyzing the axisymmetric simulation whose initial
condition has already been shown to produce a significant jet, the
dipole.  Figure~\ref{DFPevln} shows a time series of color plots
of the $\beta$ parameter overlaid with white contours depicting
poloidal field lines.  As the radial field is sheared, a toroidal
field (visible in Fig.~\ref{DFPevln} through its effect on $\beta$) is
created and amplified, leading to poloidal magnetic pressure gradients
\cite[]{De-Villiers:2003a}.  Because the initial radial field has opposite
sign on opposite sides of the equatorial plane, the toroidal field created
by the shear also changes sign across the equator.  The inner edge of
the torus begins to move inward toward the event horizon, dragging the
magnetic field with it.  As the gas moves inward, its rotation velocity
increases, further strengthening the toroidal field.  As the magnetic
pressure increases, gas is forced toward the equator where the field is
comparatively weak, bringing the regions of oppositely-directed toroidal
field closer together (see, e.g., the panel illustrating $t = 800M$).
As a result, a strong current sheet forms along the equatorial plane.

Once the inner edge of the accretion flow arrives at the black hole
($t \sim 700M$), field lines attach to the event horizon, and gas
rapidly drains off them.  Field lines slide toward the poles as they
expand outward at relativistic speed, driven by the contrast between
the high magnetic pressure near the equatorial plane and the much lower
pressure higher up along the axis.   This flow moves outward so swiftly
because it is almost entirely unencumbered by the inertia of matter.
As discussed in \cite{Hawley:2006}, a magnetic tower forms, similar
to the one predicted on analytic grounds \citep{Lynden-Bell:2003} and
seen in other simulations \citep{Kato:2004}.  However, in contrast to
these models, tightly-wrapped field lines persist only briefly; the large
vertical gradient in magnetic pressure propels the upper loops so rapidly
outward that in the steady-state the field lines are better described as
loosely helical than as a tightly-wrapped coil.  The radial field inside
the funnel has the same sign as the vertical component of the initial
dipole loop inside the initial pressure maximum; these field lines close
outside of $r=100$M, reentering the problem volume throughout the corona.

A time history of the unbound Poynting flux at $r=100M$, normalized by
the instantaneous (coordinate frame) magnetic field energy within the
bound accretion flow, is shown in Figure~\ref{2dTrtdU} for both the
three-dimensional simulation KDPg and the high-resolution axisymmetric
model.  In both cases, once the jet is established
(a process taking $\lesssim 1000M$), the Poynting flux jet persists.
It is frequently perturbed, and the amplitude undergoes fluctuations,
but its basic properties persist throughout the simulation.  Excluding
initial transients (i.e. $t<1500M$), the jet histories of the two
simulations are very similar to one another, despite the differences
in dimensionality, (poloidal)
resolution, torus configuration and equation of state: in both cases,
the ratio of jet EM luminosity to energy density in magnetic energy
is $\simeq 0.1$--0.2 per time-unit. This similarity demonstrates
that, for this topology at least, the properties of the funnel field
depend primarily on the field strength within the initial accretion flow,
and that funnel field creation is a primarily axisymmetric process.

Although the initial magnetic field had structure only on the scale
of the disk thickness, the {\it essentially laminar} dynamics of initial
inflow lead quickly to the formation of truly large-scale field.  On the
other hand, because the accretion flow takes longer to be established
than the jet does, the magnetic field in the funnel acts in effect
as a permanent boundary condition for the accretion flow.  In other
words, these simulations mimic the long-term behavior of an accretion
flow in which there is large-scale net magnetic flux threading the
event horizon and filling a cone around the axis, but no large-scale
flux piercing the disk proper.  As both the three-dimensional and
two-dimensional simulations demonstrate, it is entirely possible for
long-term quasi-stationary accretion to coexist with a strong Poynting
flux jet whose foundation is magnetic flux trapped long before.

\subsubsection{Quadrupole Topology}\label{quadrupole}

We next consider the evolution of an initial quadrupolar field in high
resolution axisymmetry.  Figure \ref{QFPevln} shows a time series of plots
illustrating $\beta$ and the poloidal field line structure for this case.
Initially, the loops of weak poloidal magnetic field enclose two regions,
located above and below the pressure maximum (see Figure \ref{topologies},
center panel). In a similar fashion to the dipole case, these loops are
sheared out, creating ever growing toroidal field and producing strong
poloidal magnetic pressure gradients. In contrast to the dipole case,
however, two current sheets are formed, one above the equatorial plane,
one below.  The magnetic pressure forces gas toward these current sheets,
where the magnetic pressure has a local minimum.

As the inner edge of the torus moves radially inward toward the black
hole, the current sheets maintain their structure.  When the field
reaches the black hole, the current sheets are located above and below
the equator.  As with the dipole, gas drains off the field lines, which
anchor themselves in the event horizon,  and the field lines between the
current sheets and the axis slide toward the poles, where the density and
pressure are very low.  These field lines expand outward relativistically,
filling the funnel region and establishing a large-scale quadrupole field.

While this field is in place it supports a Poynting-flux jet, but the power
in the jet is significantly less than that in the axisymmetric dipolar
simulation.  Averaged from $t=2800M$ to $t^\prime = 3800M$, the EM flux
in the dipolar case is twice as strong as in the quadrupolar simulation
when measured at $r=10M$, and 10 times stronger measured at $r=100M$ (we
will discuss later why the contrast was more than an order of magnitude
greater in the 3-d simulations).  Figure~\ref{TrtdUcomp} compares
the dipole jet Poynting flux with that of the quadrupole simulation as a
function of time at $r=100M$, normalized by the total magnetic field
energy in the disk.  In terms of this normalized power, the jet in
the quadrupole simulation is a factor of
2--10 weaker than that associated with dipole model; in other words,
much of the actual power contrast between the dipolar and quadrupolar
cases can be explained in terms of a reduction in this ``magnetic
efficiency".  Unlike the situation for accretion properties, contrasts
in jet luminosity from one topology to another cannot be described
solely in terms of contrasts in mean magnetic field strength.

   \cite{McKinney:2004} reported that replacing the dipole
with a quadrupole field led to a jet
that is  weaker by a factor of 2--3, but for several reasons one should
expect only qualitative, not quantitative agreement with our results.
They measured the jet luminosity at $r=40M$ and their initial pressure
maximum was much closer to the black hole ($r=12M$), so that the radial
dynamic range of their genuine accretion flow was considerably smaller.
Their disk was also geometrically thicker, partly because their code
solved a total energy equation, and partly because their initial state
was hotter.

Simple geometry accounts for the qualitative contrast between the jet
produced by an initially quadrupolar magnetic geometry and that driven
by an initially dipolar field.  In the quadrupolar case, relatively small
amplitude vertical oscillations in the inner region of the accretion flow
can lead to reconnection between magnetic field in the funnel and magnetic
field in the disk because vertical field of both signs is available in
the same hemisphere.  The result is to break the magnetic connection
between the black hole horizon and large radius.  Put another way, the
presence of closed field loops on the same side of the equator offers
the opportunity for easy collapse of these loops to zero.  In the dipole
case, however, the sign of the vertical magnetic field  passing through
the equator is the same everywhere inside $r=25M$.

The net result of this geometric contrast is that, while quadrupolar
field configurations can pump magnetic field into the outflow funnel and
support a jet, they can also equally easily destroy such a jet by reconnection
with oppositely-directed field.  Jets in this case are consequently very
unsteady, and exhibit strong variability in the Poynting flux, as shown
in Figure~\ref{TrtdUcomp}.  This variability is a major factor in
depressing the time-averaged EM luminosity of jets driven by quadrupolar
fields.

Because the code simulates the equations of ideal MHD, reconnection is
numerical and occurs at the grid scale. As such, the reconnection rate
is resolution dependent (lower resolution promotes reconnection). We
cannot therefore make specific predictions regarding the variability
characteristics of jets in Nature, as this is determined (in our
simulations at least) by competition between processes that are
physical in character (e.g., orbital dynamics, magnetic forces) and
purely numerical (the reconnection rate).

The rate of reconnection is also determined by details of the field
geometry, in addition to its basic topology.  To investigate this sort
of dependence, we have conducted a series of lower resolution ($256^2$)
axisymmetric quadrupolar simulations in which we varied the size of the
initial torus, and hence the scale of the initial field loops.  We kept
the inner edge of the torus fixed at $r=15M$, but reduced the angular
momentum $\ell$ in order to reduce the vertical thickness of the
torus, and consequently reduce the vertical extent of the quadrupolar
loops. Simulations were carried out for a variety of values of $\ell$,
ranging from $\ell = 4.9$ to $\ell=5.2$ in steps of $\ell=0.05$.

For the purposes of our discussion here, we focus on the $\ell=4.9$
torus, which serves to illustrate the impact of vertical scale height on
the evolution of the quadrupole topology. Specifically, setting $\ell =
4.9$ results in a torus which is initially a factor of two thinner (in
terms of the density scale height, measured along surfaces of constant
radius at the location of the pressure maximum) than the $\ell = 5.2$
torus previously discussed. The vertical thickness of the resulting
quadrupolar field loops is then very similar to that of loops in
the sub-Keplerian torus used in QDPa.

The initial evolution of this system is shown in Figure \ref{QNPevln}. The
evolution proceeds as before: current sheets form above and below
the equatorial plane as the inner edge of the torus moves toward the
black hole.  This time, however, the two current sheets are much closer to
the equatorial plane (contrast Fig.~\ref{QNPevln} with Fig.~\ref{QFPevln}).
Around $t=800M$ the current sheets buckle,
apparently due to an instability, and the coherent poloidal field
structure is destroyed.  A similar phenomenon was observed in QDPa,
which had a similar scale-height:
the current sheets destabilized far from the black hole, destroying the
coherent poloidal field structure. As a result, no large scale poloidal
field was dragged into the near horizon region by the torus, and the
resulting field inside the funnel region is substantially reduced.
We note that in three-dimensions the presence of non-axisymmetric MRI
modes are likely to enhance the rate of reconnection, further decreasing
the likelihood that an organized quadrupole configuration will survive
to create a significant funnel field.

These additional simulations show that the strength of Poynting-dominated
jets depends on field geometry in two senses: it is topology-dependent
and shape-dependent.  Quadrupolar magnetic topology leads to easier
reconnection, and therefore weaker funnel fields and jets, than does
dipolar topology, and geometrically thinner quadrupole loops make
reconnection still more rapid.

\subsubsection{Multiple Loop Topology}\label{multipleloop}

We have shown how large dipolar loops can create strong jets, but what
happens if the oppositely-directed vertical field on the far side
of the loop is accreted?  This did not occur in KDPg, in part because
it did not run long enough and in part because the finite mass in our
torus meant that the matter outside the pressure maximum moves outward
over time, not inward.   In a real accretion disk, however, continuing
mass re-supply will force the far sides of even large dipolar loops
to accrete.

To explore this sort of event, we ran a two-dimensional
simulation whose initial state contained a series of four narrow dipolar
loops (described in \S\ref{numerics}). The initial evolution of
this topology is shown in Figures \ref{MFPevln1} and \ref{MFPevln2}.
The first of these figures shows the descent of the innermost radial
loop into the near-horizon region and the establishment of a Poynting
flux jet. The second figure shows the descent of the second (moving out
in radius) of these loops and its interaction with the field structures
established by the first loop. 

Over the timescale considered, the evolution of this model is determined
by the two innermost loops, both initially inside the pressure maximum
(see Fig.~\ref{topologies}).  As for a single dipole loop, the first loop
forms a funnel dipole field by $t=700M$.  However, its shape differs
significantly from the single loop because the second loop begins its
descent toward the inner disk even before the inner edge of the first
loop nears the event horizon (Fig.~\ref{MFPevln1}).  As the second loop
moves inward, it displaces the field lines of the first loop, forcing
the midplane portions of the outer half of the first loop inward, and
the off-plane portions of the first loop to higher altitude at their
original radius.  The result is that the first loop is very nearly
transformed into a pair of loops, one on each side of the midplane.
After arriving in the inner disk, the field of the second loop expands
rapidly toward the funnel region.  It pushes on the almost closed
sections of the first loop, triggering rapid reconnection where its
oppositely-directed segments pass very close to one another at small
radius (Fig.~\ref{MFPevln2}).  At this point, the transformation of the
first loop into a disconnected quadrupolar loop pair is complete, and the
loops fly off to large distance, emptying the funnel of magnetic field.
The funnel is then refilled with flux from the second
loop.  These dramatic events are mirrored in the very large fluctuations
in normalized Poynting luminosity shown
in Figure~\ref{TrtdUcomp}.  Jumps in output even greater in relative
terms than those displayed by the quadrupolar simulation are frequent,
and can be associated with these episodes of funnel field establishment
and annihilation.  Note that while the multiple-loop jet is highly
variable, similar to the quadrupolar jet, the instantaneous
strength of the Poynting-flux jet (when it is strong) is similar to
that of model KDPg when normalized for disk magnetic field strength.
In this sense, the ``efficiency" of these systems in using magnetic
field generated in the disk for jets appears to be a function of
poloidal field character, declining as the relative importance
of higher multipole moments increases.

In interpreting this simulation, it is important to bear in mind that,
unlike any of the others reported in this paper, its accretion flow never
achieves anything resembling a quasi-stationary state.  The multiple
loop structure also requires high resolution to prevent reconnection from
occurring at too rapid a rate.  It is for this reason that, unlike
the large dipolar loop and quadrupolar caess, we report no 3-d
simulation results for this field geometry.  Thus, we can be less
confident than
in the other cases that the specific phenomenology we observe may
be associated with real systems in a state of continuing accretion.
What it {\it does} demonstrate, however, is that accretion of a series
of closed dipolar loops can lead to a sequence of field establishment
and destruction rather similar to what was seen in the quadrupolar case,
and through mechanisms that also resemble those seen when the initial
field is quadrupolar.  It also supports our qualitative inference that
maintenance of a consistent sign of vertical field is essential to
the long-term support of a Poynting-dominated jet.  The lifetime for
an episode of strong outflow is roughly the inflow time for the outer
anchor point of the loop.

\subsubsection{Toroidal Field Topology}\label{toroidal}

In the poloidal cases, magnetic field is injected into the funnel by
strong toroidal field pressure in the plunging region.  One might expect
that this process could act equally efficiently when the magnetic field
is toroidal from the start.  If the toroidal field has a consistent sign,
the radial component created when it expands outward would also have a
consistent sign, satisfying the coherent vertical field criterion we
have inferred from study of the poloidal simulations.  If so, the toroidal
field case should also drive a powerful jet.  However, it does
nothing of the sort.

Figures~\ref{TDPevln1} -- \ref{TDPevln2} are a series of snapshots of
the $\beta$ parameter and the radial field structure in
the evolving torus of simulation TDPa.  In contrast to the poloidal
loop simulations, there is no initial radial field to generate toroidal
field through shear.  Instead, the evolution is governed entirely by
the development of the nonaxisymmetric MRI.  The fastest-growing modes
for the MRI in this context have large $k_z$ and azimuthal wavelength
$\sim v_A/\Omega$, which is much shorter than the disk thickness when the
field is weak \citep{Balbus:1992}.  The magnetic structures that develop
are then predominantly on short lengthscales.  Note the rapid spatial
variation in sign in the radial field in Figure~\ref{TDPevln2}.  With
time, the radial field features become somewhat more radially elongated,
but field lines do not connect extended regions;  no large-scale coherent
poloidal field is generated from the initial toroidal field.  Magnetic
field amplification is less than in the poloidal field initial conditions,
and $\beta$ never becomes less than 1 near the hole.  Matter tends to
remain with the field rather than drain either into the hole or toward
a localized region such as the equator, as seen in the dipole case.
Any vertical expansion of the field would therefore have to overcome the
inertia of the matter and, as can be seen from Figure \ref{TDPevln2},
this does not occur.

The absence of a funnel-field in the toroidal topology follows directly
from the specific properties of the toroidal field MRI: the poloidal
fields created by the MRI are typically small scale.  No radially-extended
poloidal field connects the near horizon region to the disk.  What radial
field there is is accreted faster than it can tap into the differential
rotation and generate strong near-hole fields.  The main disk-body in
the toroidal case remains gas pressure-dominated at all times, and there
are no strong magnetic pressure gradients to drive an expansion of the
field into the funnel region.  The absence of a funnel field means there
is no Poynting flux jet.

It appears, then, that creation of a strong funnel field, and therefore
a strong Poynting jet when the black hole spins rapidly, depends
essentially on the existence of some coherent poloidal field in the
initial state.  Purely toroidal field initial conditions can lead to
short coherence-length poloidal loops with rapidly alternating sign,
but no large-scale poloidal field components.

\section{Summary and Conclusions}\label{conclusion}

Through a series of numerical studies, we have examined the influence
of the initial field topology on the properties of magnetized black hole
accretion flows. In particular, we have performed full three-dimensional
simulations of three different initial field topologies: a large-scale
dipolar loop, a pair of quadrupolar loops separated by the equatorial
plane, and a purely toroidal field.  Apart from the imposed initial field
topologies, the initial conditions used in each of these simulations
were identical: a $\Gamma = 5/3$ hydrostatic torus with a slightly
sub-Keplerian distribution of angular momentum around a rotating
($a/M=0.9$) black hole.  The two poloidal field topologies were evolved
to $t=10000$M, while the toroidal case was evolved to $t=29000$M.

\begin{figure*}
\begin{center}
\leavevmode
\includegraphics[width=0.48\textwidth]{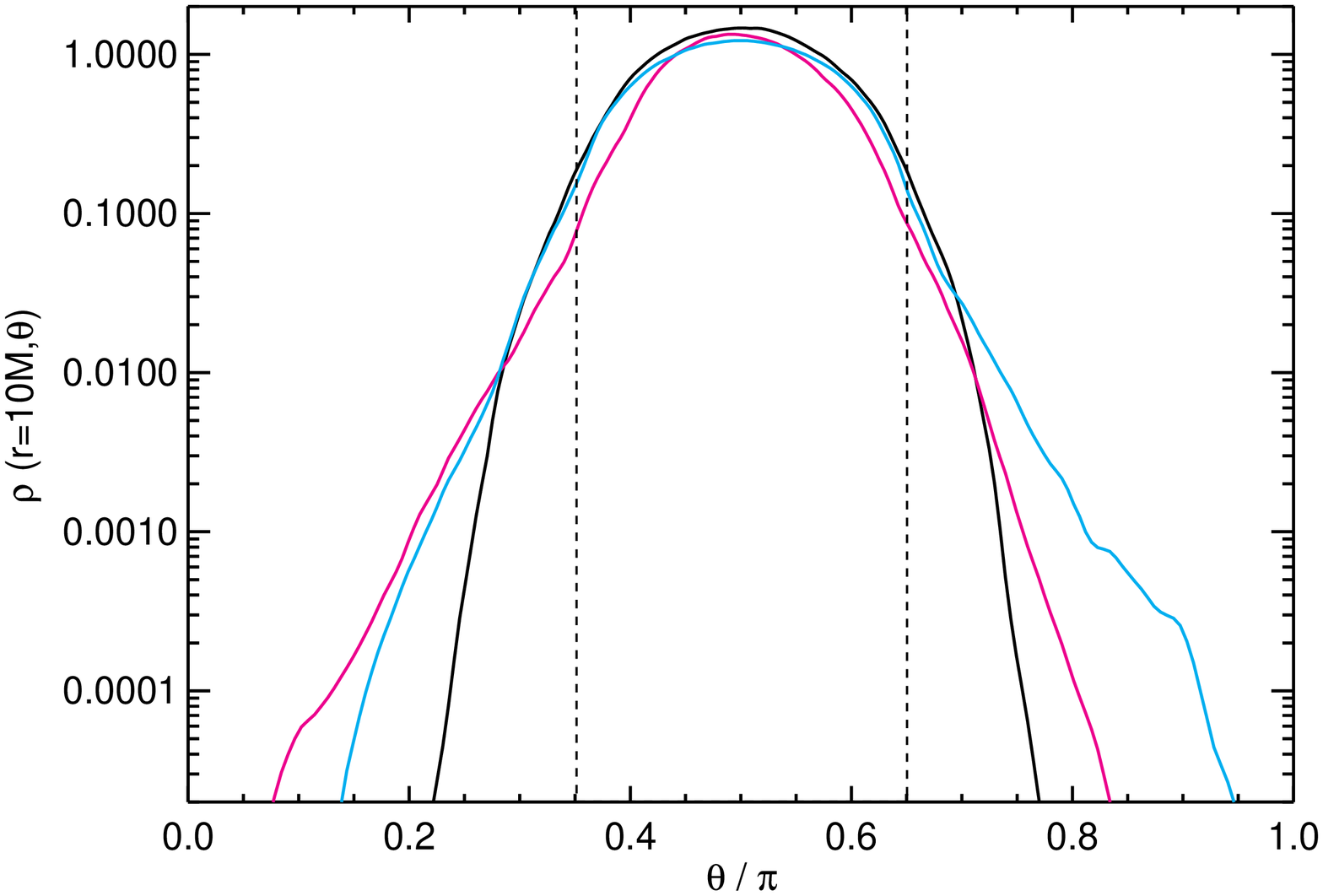}
\includegraphics[width=0.48\textwidth]{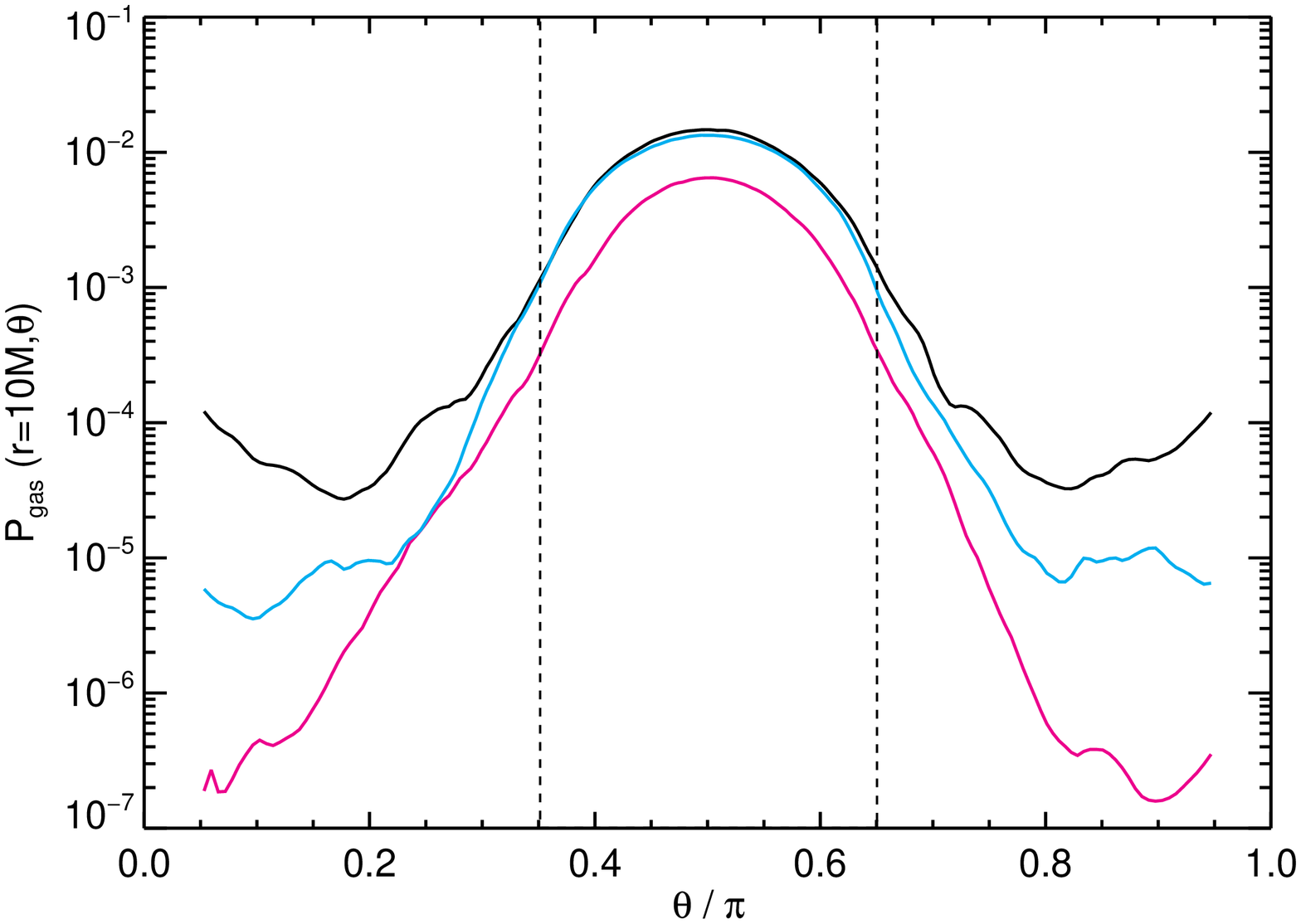}
\includegraphics[width=0.48\textwidth]{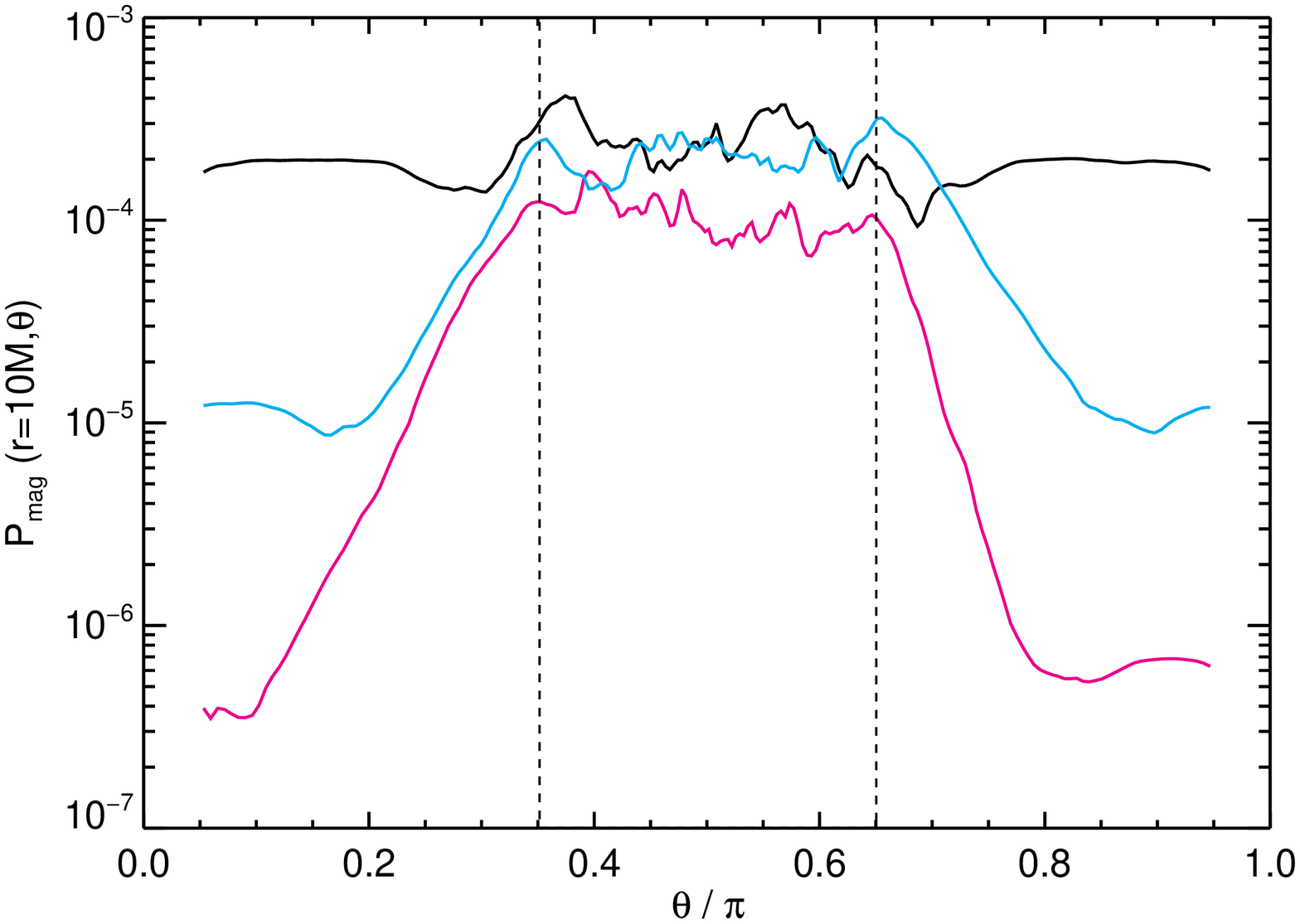}
\includegraphics[width=0.48\textwidth]{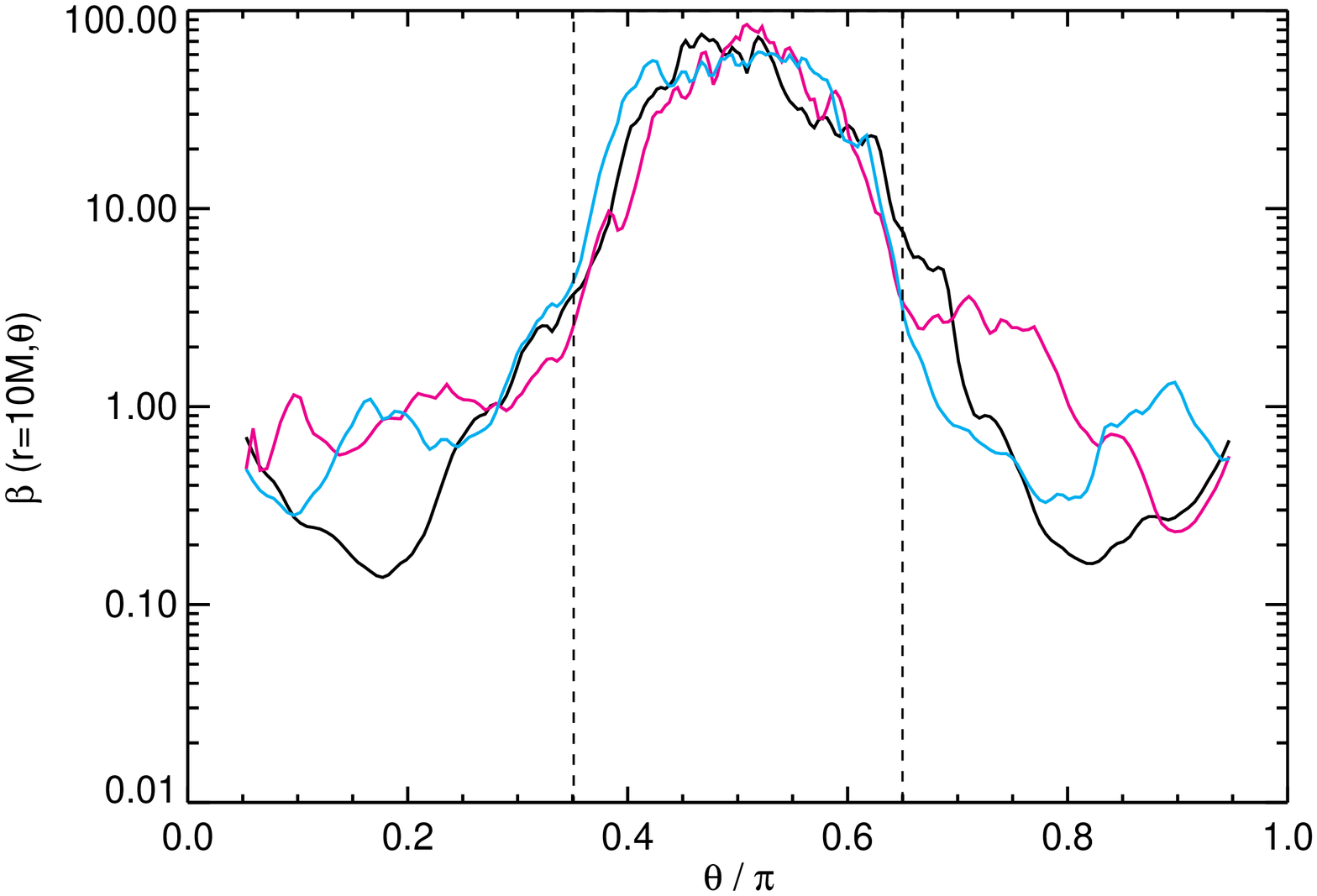}
\end{center}
\caption[]{Time-averaged angular profiles of density $\langle \rho (\theta; r=10M) \rangle_{A}$, gas
pressure $\langle P (\theta; r=10M) \rangle_{A}$, magnetic pressure $\langle \frac{1}{2} ||b||^{2} (\theta; r=10M) \rangle_{A}$ and $\beta$ parameter $\langle 2 P / ||b||^{2} (\theta; r=10M) \rangle_{A}$ for simulations
KDPg (black lines), QDPa (blue lines) and TDPa (magenta lines).
Dotted lines denote the (approximate) boundary of the disk and corona.
Note that in the case of TDPa and QDPa, the region of bound matter 
extends all the way to the axial cutout. KDPg and QDPa are averaged over 8000--$10000 M$ and TDPa is
averaged over 2--$2.2\times 10^4 M.$}
\label{angcor} 
\end{figure*}

\begin{figure*}
\begin{center}
\leavevmode
\includegraphics[width=0.48\textwidth]{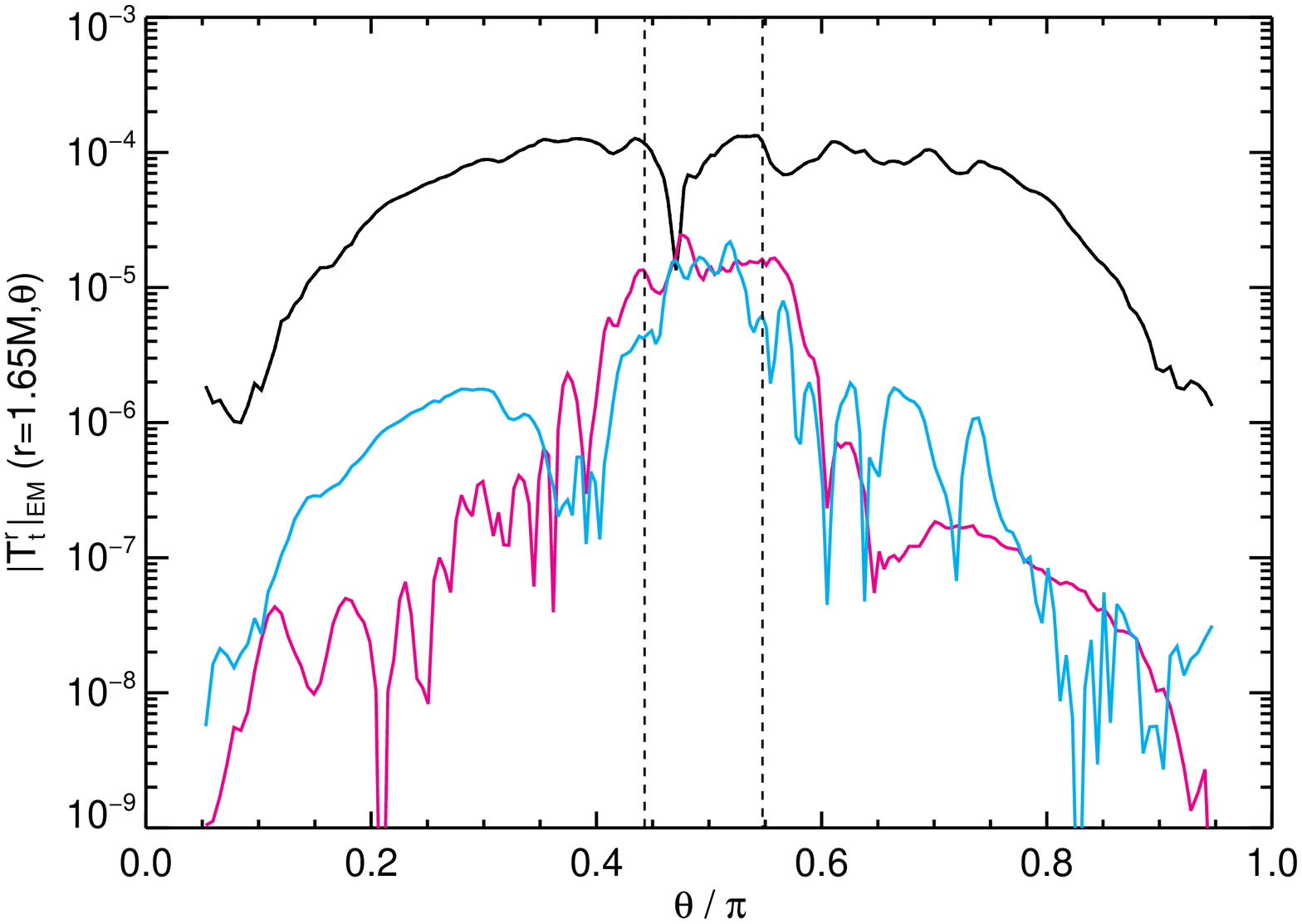}
\includegraphics[width=0.48\textwidth]{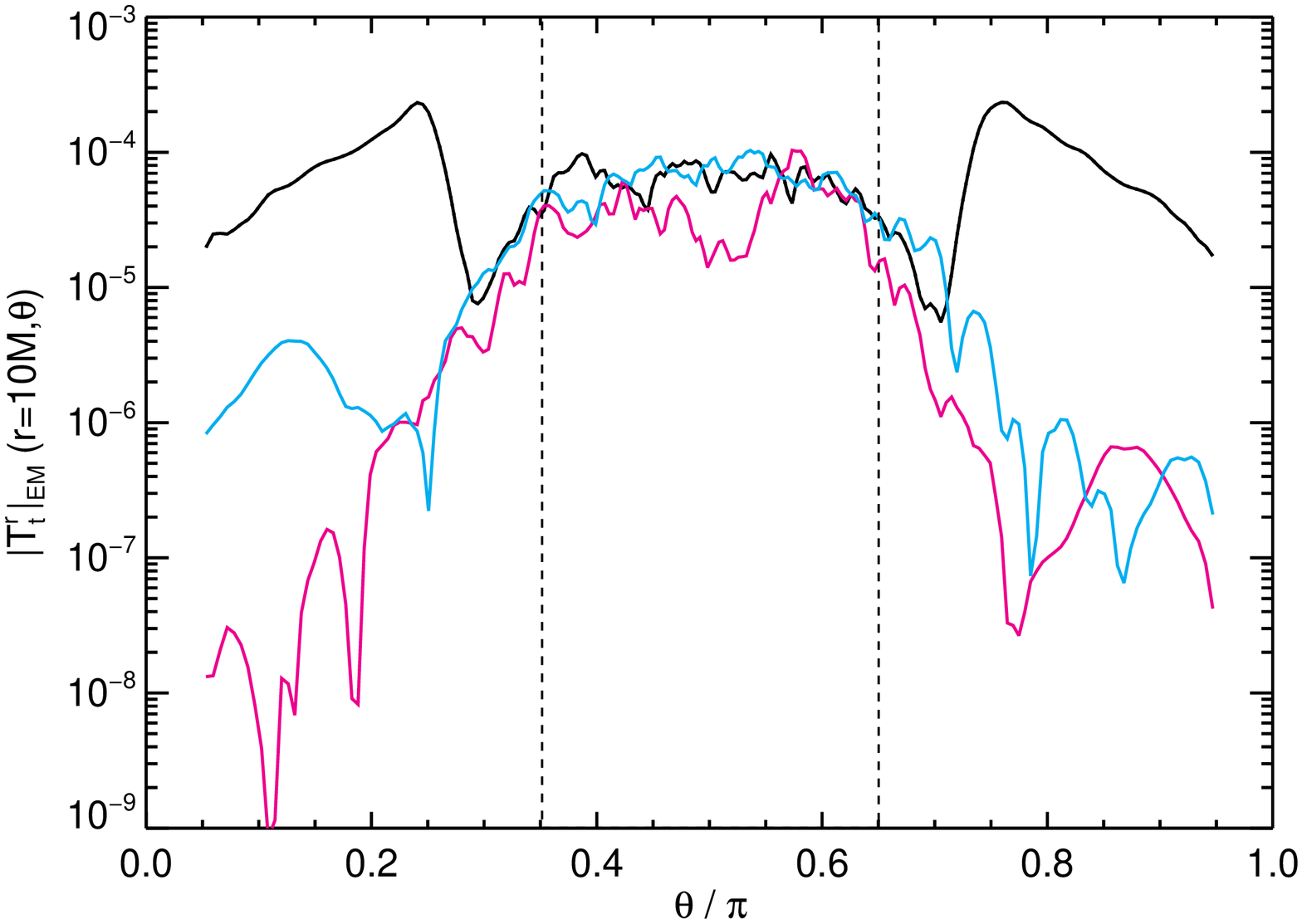}
\includegraphics[width=0.48\textwidth]{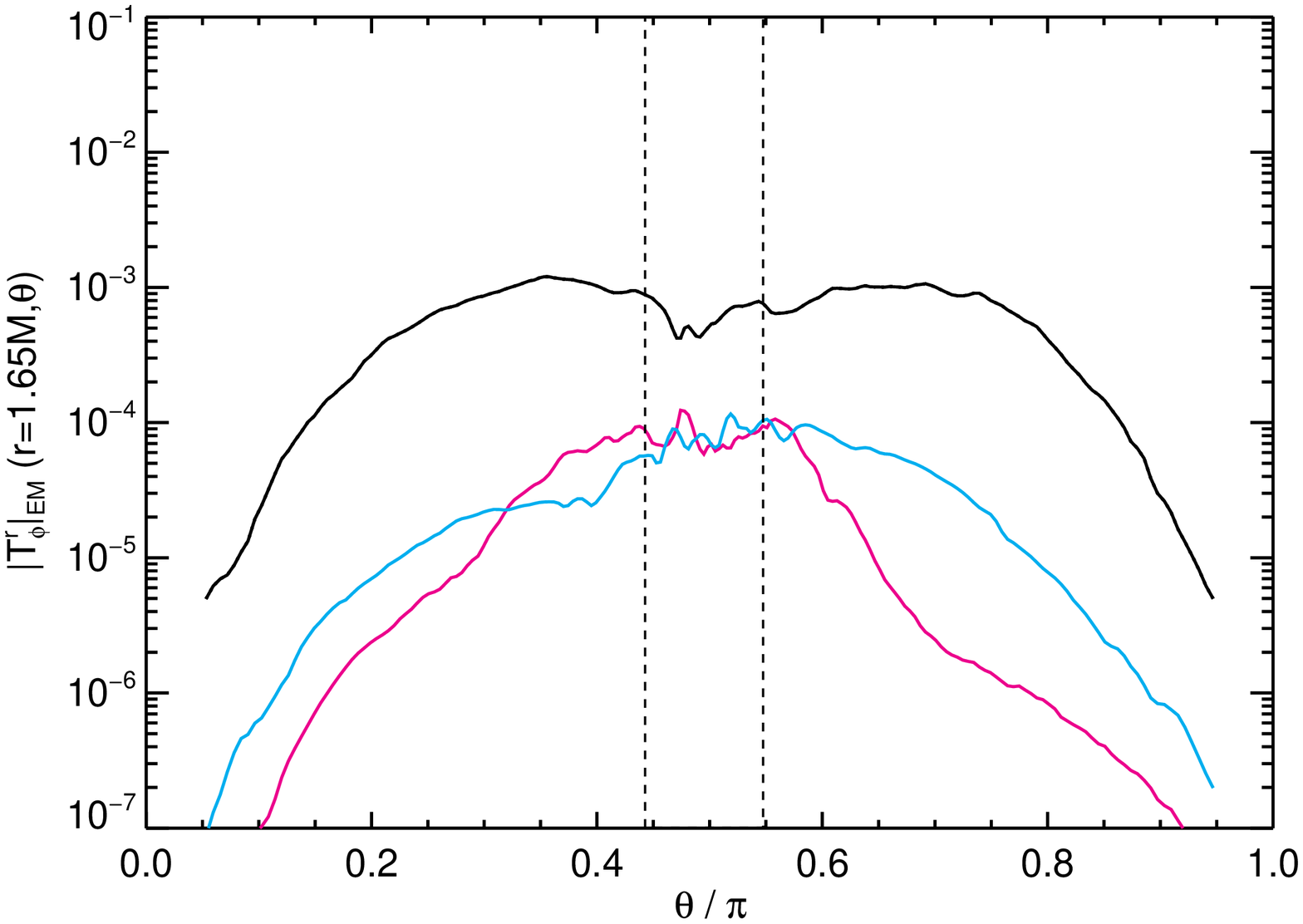}
\includegraphics[width=0.48\textwidth]{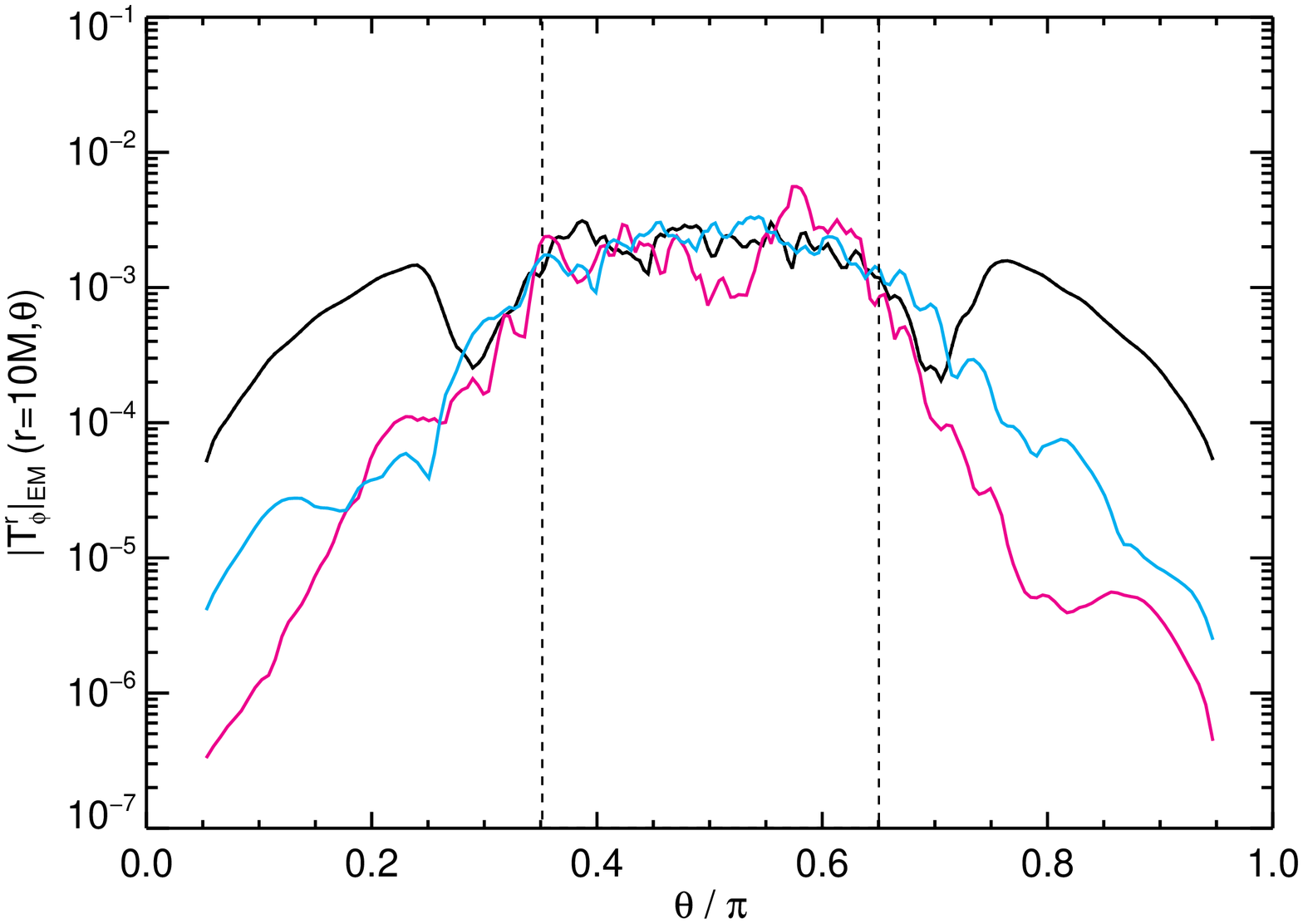}
\end{center}
\caption[]{Time-averaged angular profiles of $\langle |T^r_{t} |_{\; \mathrm{(EM)}} (\theta; r) \rangle_{A}$ (top row) and $\langle |T^r_{\phi} |_{\; \mathrm{(EM)}} (\theta; r) \rangle_{A}$ (bottom row) for
simulations KDPg (black lines), QDPa (blue lines) and TDPa (purple
lines) at $r=1.65$M (left column) and $r=10$M (right column). Dashed
lines denote the (approximate) boundary between the disk body and
the corona. KDPg and QDPa are averaged over 8000--$10000 M$ and TDPa is
averaged over 2--$2.2\times 10^4 M.$
}
\label{angtrptrt} 
\end{figure*}

\begin{figure*}
\begin{center}
\leavevmode
\includegraphics[width=0.48\textwidth]{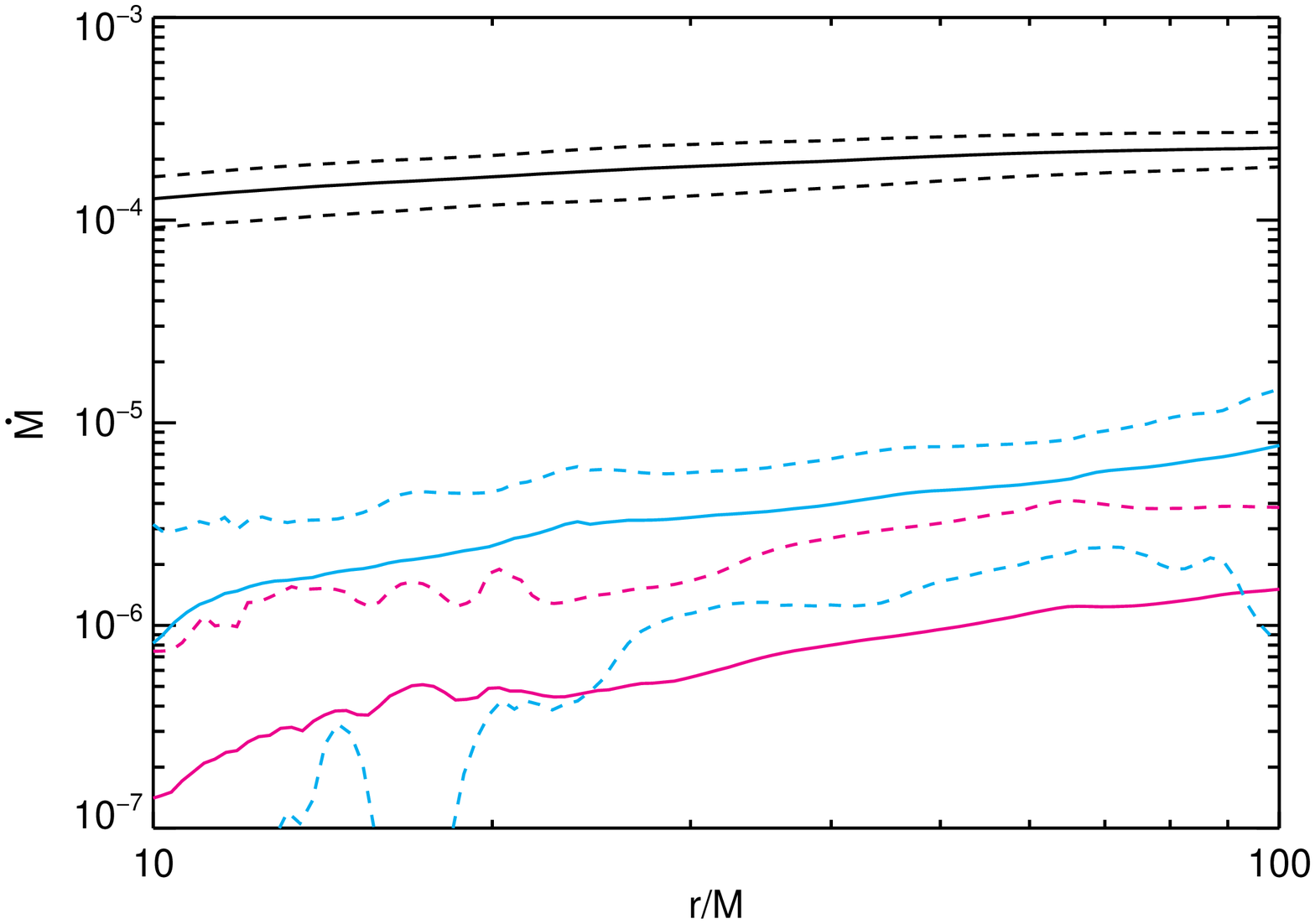}
\includegraphics[width=0.48\textwidth]{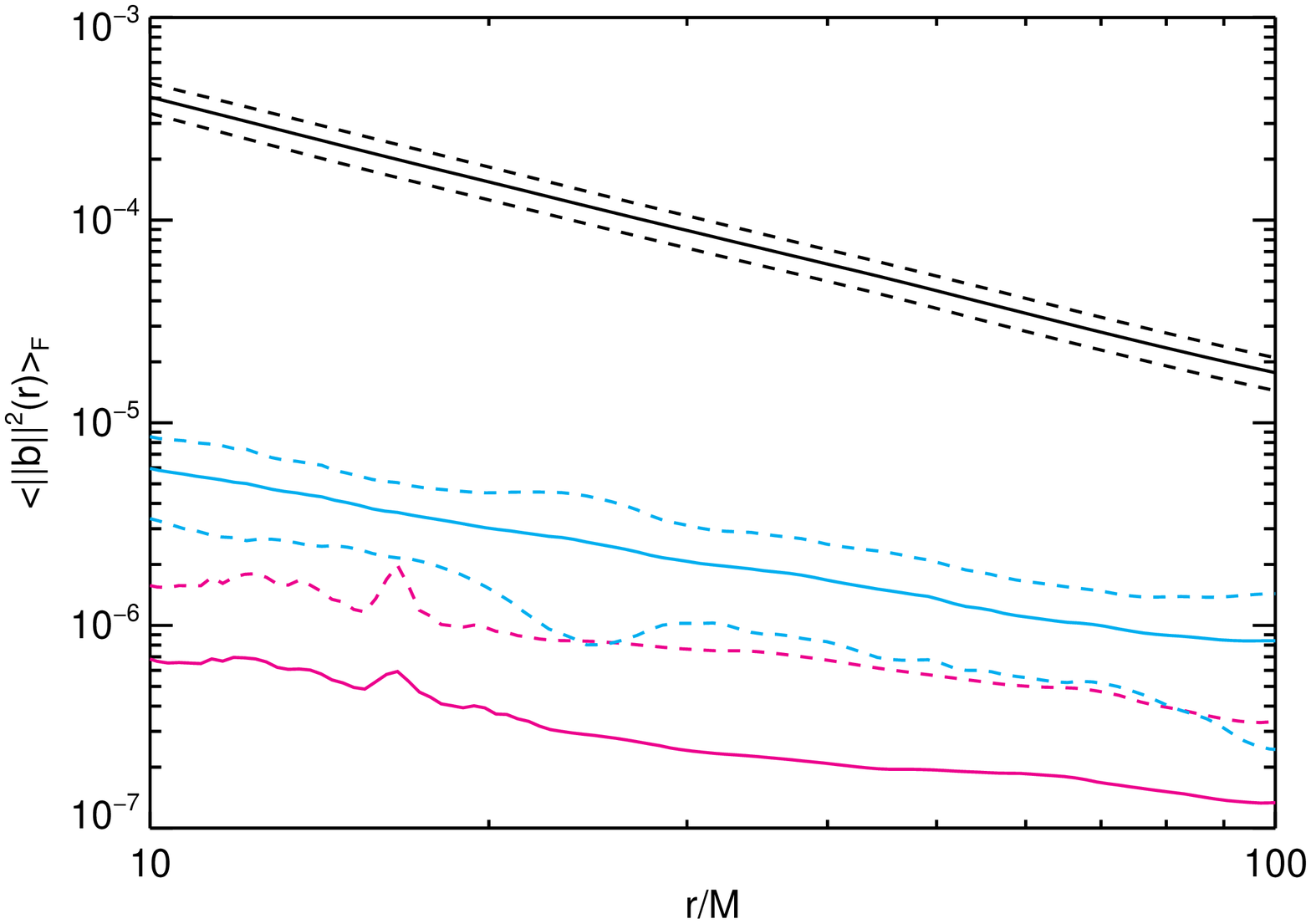}
\includegraphics[width=0.48\textwidth]{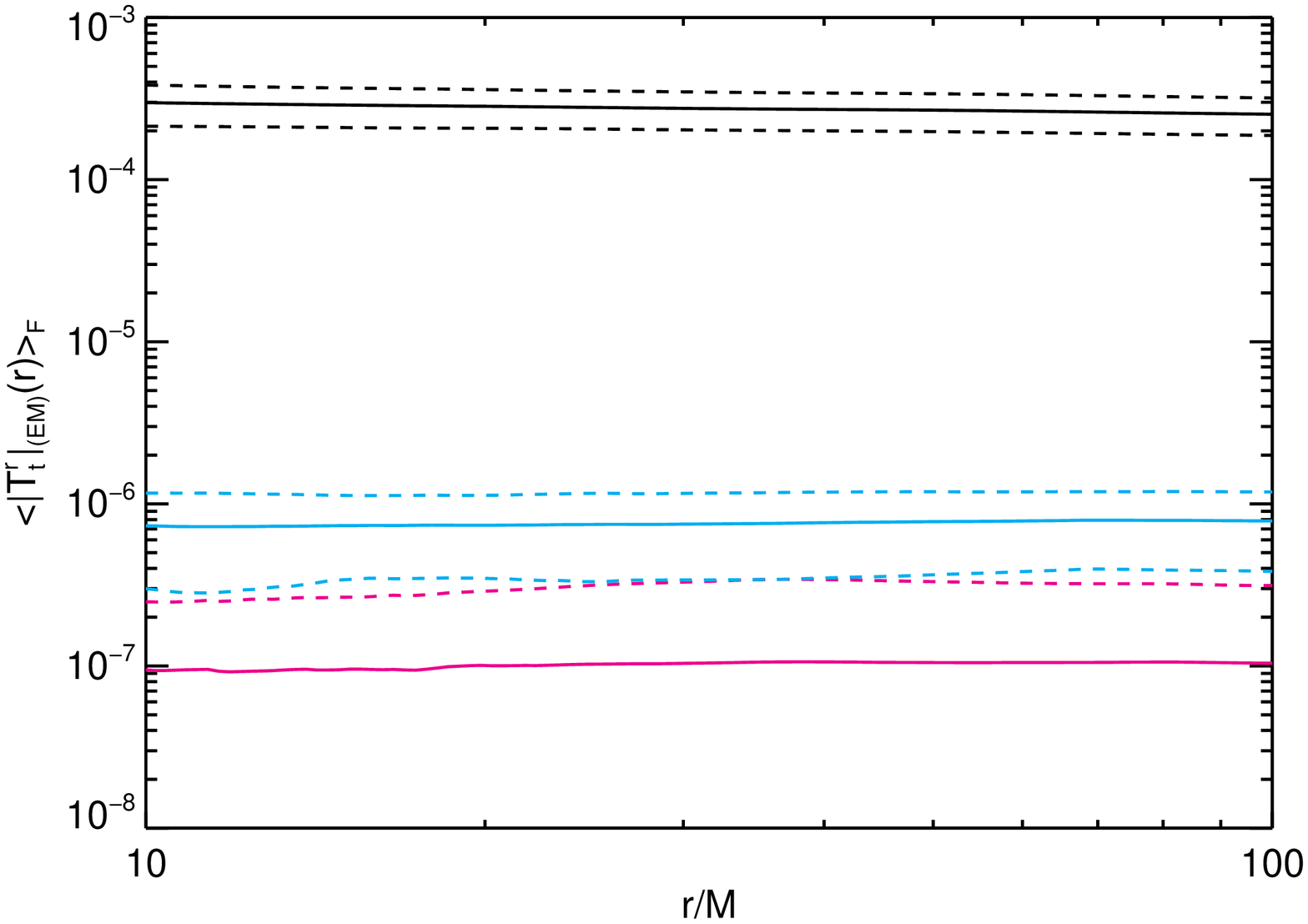}
\includegraphics[width=0.48\textwidth]{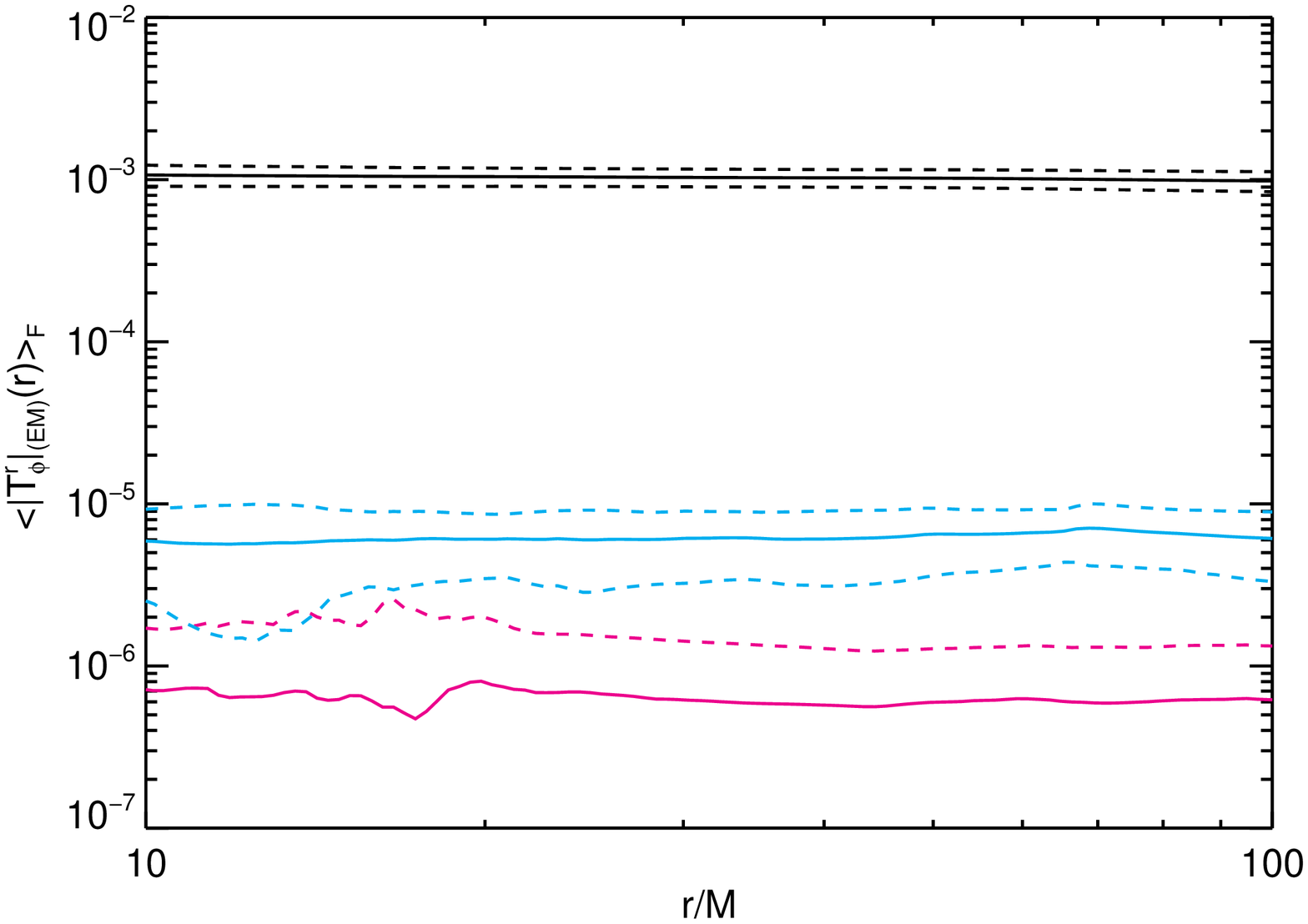}
\end{center}

\caption[]{Time-average shell integrals of data from unbound outflows as a function of radius for models KDPg (black lines), QDPa (blue lines), and TDPa (magenta lines).
Shown are mass outflow rate $\dot{M}$ (top left panel), $\langle ||b||^2 (r) \rangle_{F}$
(top right panel); along with the electromagnetic energy flux, $\langle |T^r_{t} |_{\; \mathrm{(EM)}} (r) \rangle_{F}$
(bottom left panel) and angular momentum flux $\langle |T^r_{\phi} |_{\; \mathrm{(EM)}} (r) \rangle_{F}$ (bottom right panel).  KDPg and QDPa are time-averaged over $4000-10000$M while TDPa is averaged over $12500-18500$M.  Dashed lines show $\pm1$ standard deviation from the average.}

\label{ubndavgplt1} 
\end{figure*}

\begin{figure*}
\begin{center}
\leavevmode
\includegraphics{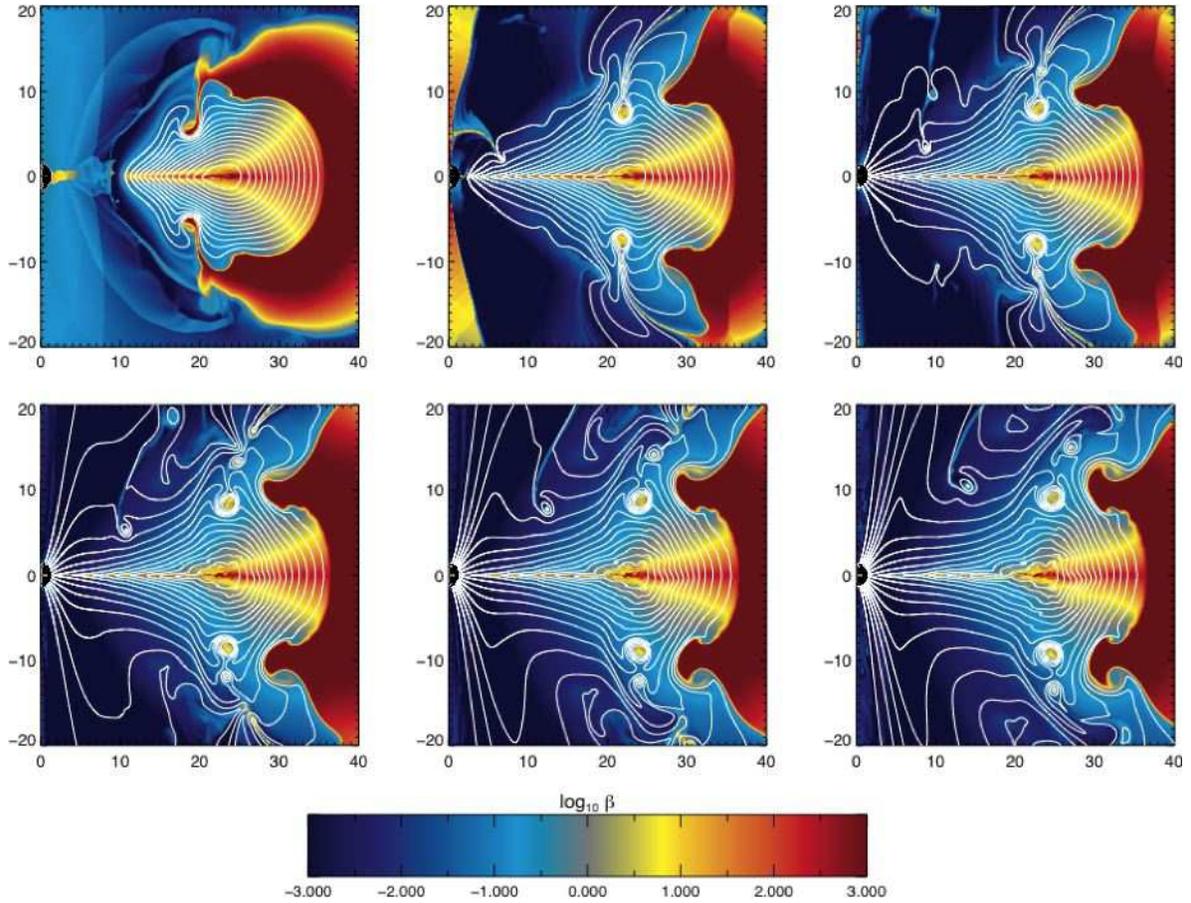}
\end{center}
\caption[]{Evolution of the magnetic field (white contours) and gas
$\beta$ parameter (filled contours) at $t = 500,700,750,800,850,900$M
for the high-resolution, axisymmetric simulation of the dipole field
topology. Dark red denotes regions of the simulation that are gas
pressure dominated and dark blue denotes regions that are magnetic
field dominated.}
\label{DFPevln} 
\end{figure*}

\begin{figure}
\begin{center}
\leavevmode
\includegraphics[width=0.5\textwidth]{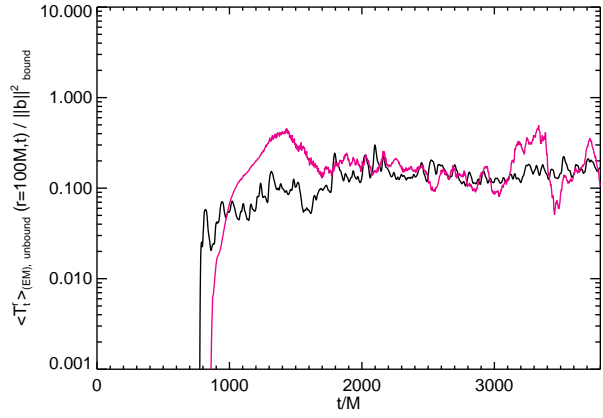}
\end{center}
\caption[]{Time history of the Poynting flux crossing the $r=100M$
surface normalized to the total magnetic field strength within the disk
for simulation KDPg (black lines) and the high resolution axisymmetric
simulation of the dipole (magenta line).}
\label{2dTrtdU} 
\end{figure}

\begin{figure*}
\begin{center}
\leavevmode
\includegraphics{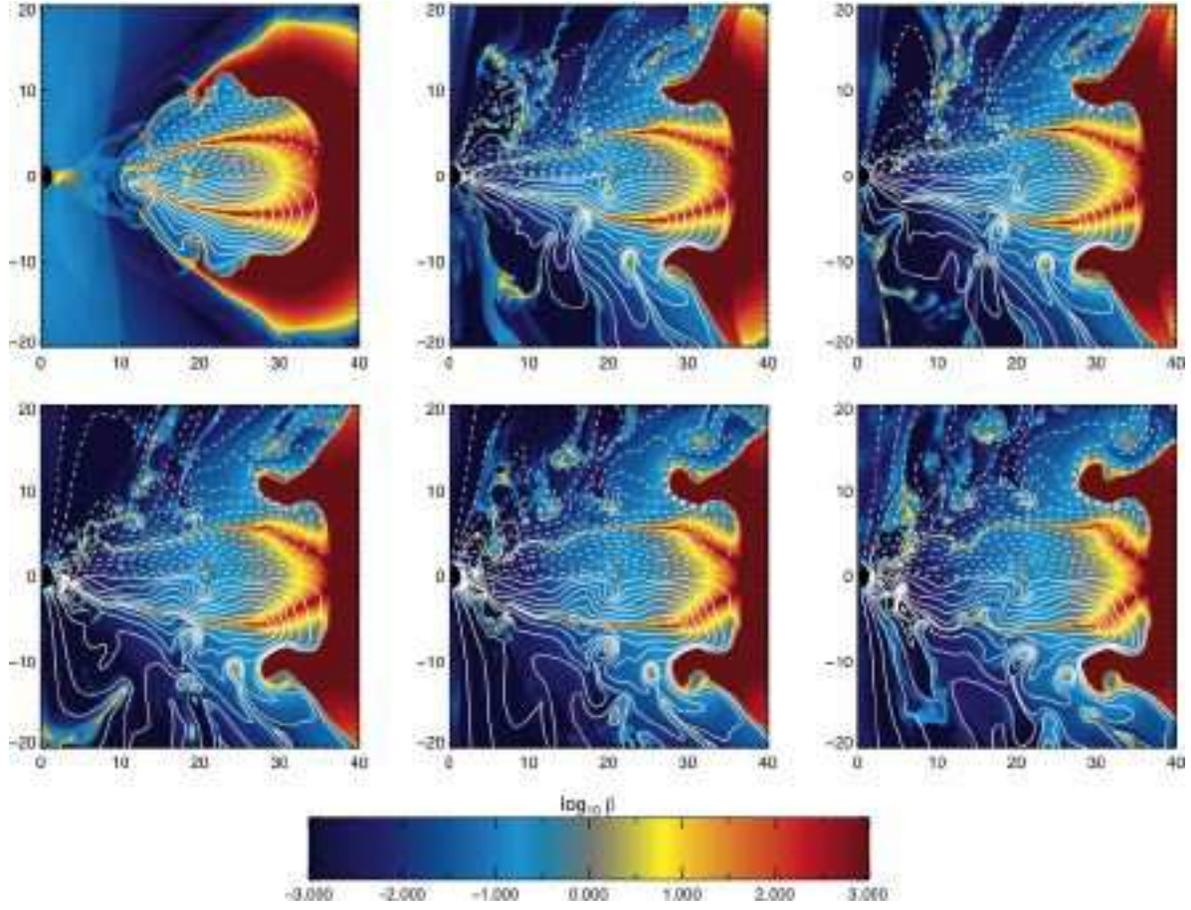}
\end{center}
\caption[]{Field evolution in the high-resolution,
axisymmetric simulation of the quadrupole field topology. Solid white
contours denote field lines that have the same parity as the dipole,
dashed contours show field lines with the opposite parity. Times
shown are the same as in Figure~\ref{DFPevln}.}
\label{QFPevln} 
\end{figure*}

\begin{figure}
\begin{center}
\leavevmode
\includegraphics[width=0.5\textwidth]{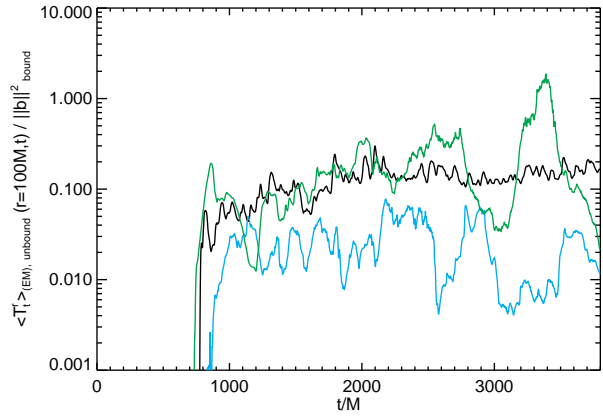}
\end{center}
\caption[]{Time
history of the Poynting flux crossing the $r=100M$ surface 
normalized by the total magnetic field strength within the disk.
The KDPg dipole simulation is shown in black , the quadrupole
simulation in blue, and the multiple loop simulation is green.}
\label{TrtdUcomp} 
\end{figure}

\begin{figure*}
\begin{center}
\leavevmode
\includegraphics{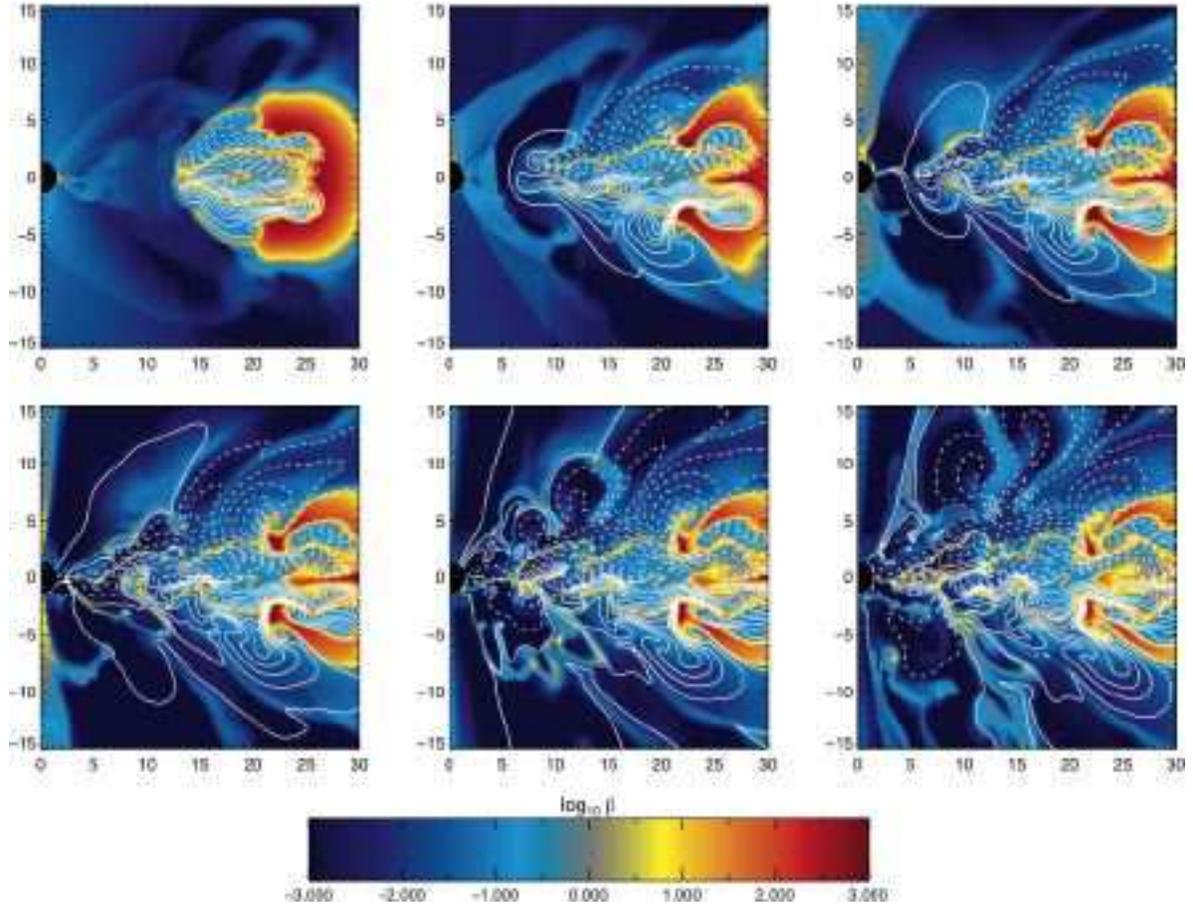}
\end{center}
\caption[]{As in Figure \ref{QFPevln} for the quadrupole field topology 
overlaid on the narrow $\ell=4.9$ constant angular momentum torus.}
\label{QNPevln} 
\end{figure*}

\begin{figure*}
\begin{center}
\leavevmode
\includegraphics{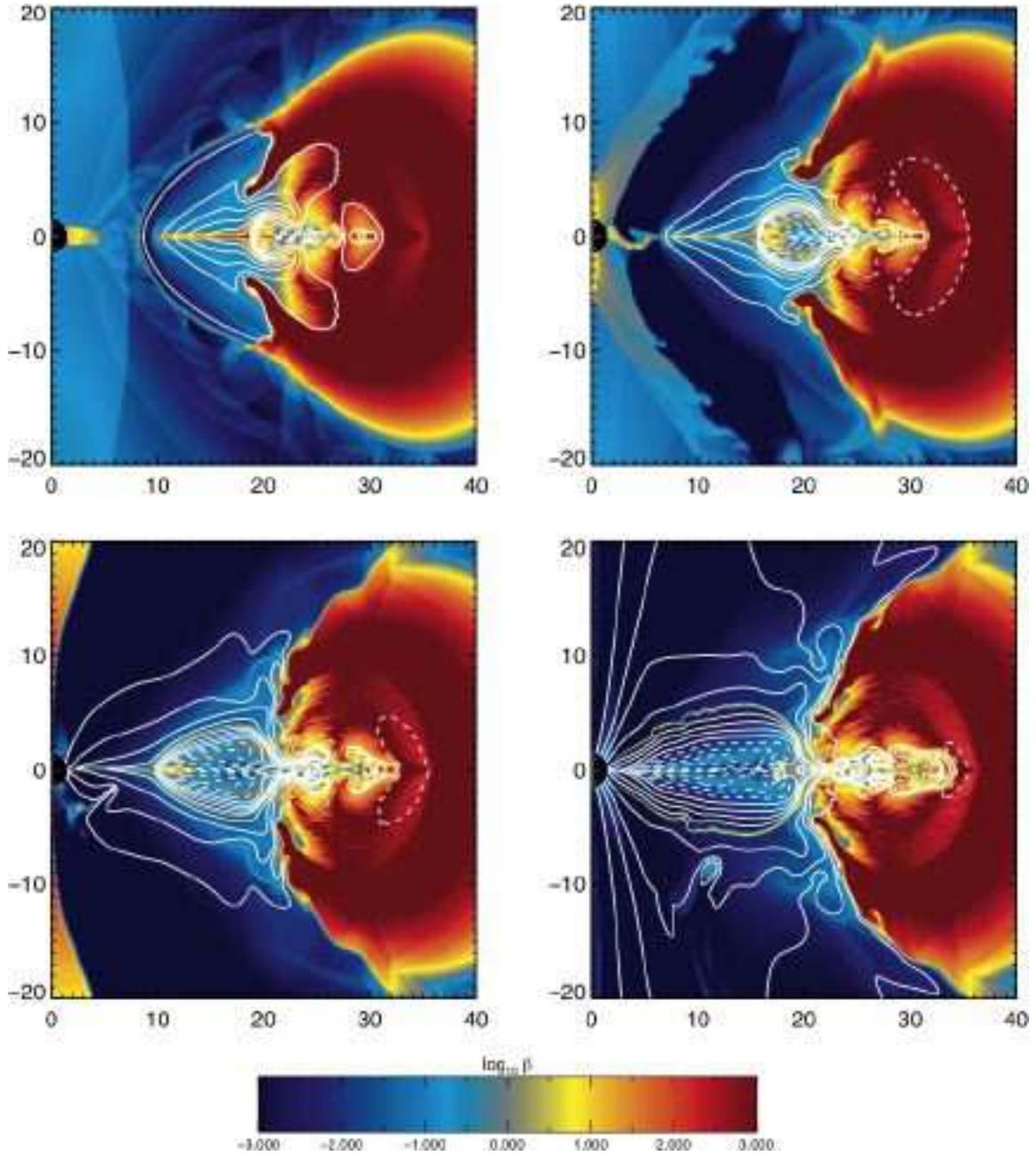}
\end{center}

\caption[]{ Axisymmetric simulation of a series of four narrow dipolar
loops at $t = 400,500,600,700$M, showing the initial infall of the
innermost loop and subsequent formation of a large-scale dipole
field. Solid white contours denote field loops of the same polarity
as used in the dipole and quadrupole topologies, dashed white contours
denote loops of opposite polarity.}

\label{MFPevln1} 
\end{figure*}

\begin{figure*}
\begin{center}
\leavevmode
\includegraphics{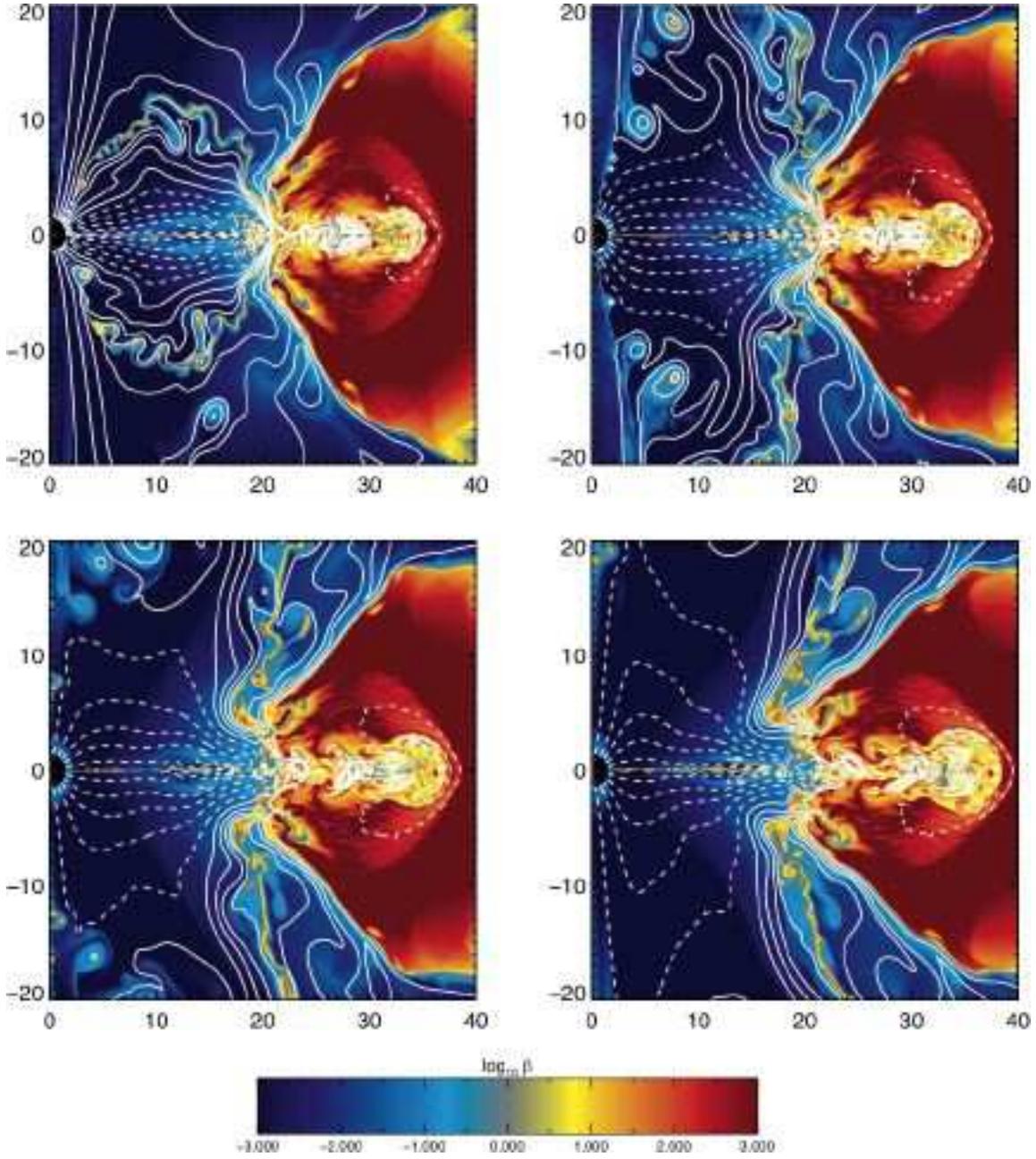}
\end{center}
\caption[]{As in Figure \ref{MFPevln1} at $t =
750,800,850,900$M, showing the interaction of the second
field loop with the dipole field established by the innermost
field loop.  The initial dipole field is destroyed but a new
large-scale field is subsequently established.}
\label{MFPevln2} 
\end{figure*}

\begin{figure*}
\begin{center}
\leavevmode
\includegraphics{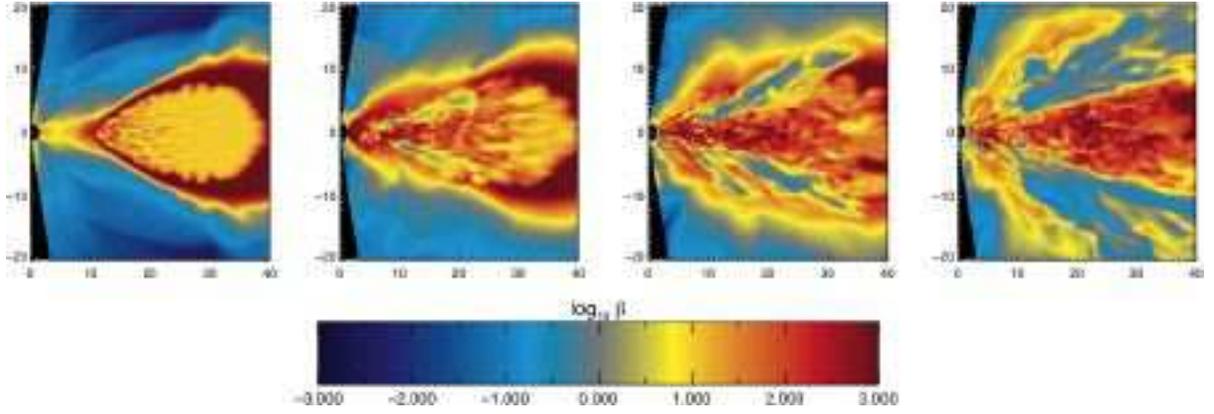}
\end{center}
\caption[]{Azimuthally averaged gas $\beta$ parameter in the toroidal
field simulation at (from left to right) $t=2000,4000,5600,7200$M. The
equatorial region remains gas pressure dominated throughout the initial
evolution, in marked contrast to all of the poloidal topologies.}
\label{TDPevln1} 
\end{figure*}

\begin{figure*}
\begin{center}
\leavevmode
\includegraphics{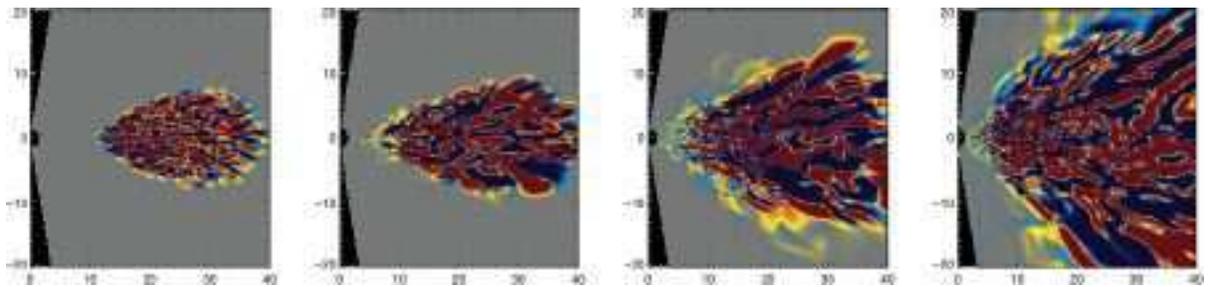}
\end{center}
\caption[]{Poloidal structure of ${\cal B}^r$ for $\phi = 0.25\pi$ in the toroidal
field simulation at (from left to right) $t=2000,4000,5600,7200$M. Red denotes strong, positive field
strength, blue strong negative field strength. Grey denotes zero field.}
\label{TDPevln2} 
\end{figure*}

\cite{De-Villiers:2003b} described the late-time structure of the
resulting accretion flow in terms of distinct components: the main
disk body, the coronal envelope, and an unbound outflow which
itself could be separated into a matter-dominated funnel-wall jet
and a Poynting-flux jet in the funnel interior.  We have found that
the late-time averaged properties of the main disk body exhibit only
modest quantitative contrasts attributable to different initial field
topologies: the dipolar and quadrupolar cases are roughly similar, and
both create somewhat stronger magnetic fields than the toroidal case.
So long as there is some weak magnetic field present, the MRI generates
turbulence and drives accretion, although the character of the purely
toroidal MRI is somewhat different from that produced by poloidal fields.
Because the Maxwell stress controls the surface density required to
support a given accretion rate, and the cascade of energy to smaller
scales in the MHD turbulence leads ultimately to disk heating, we expect
that the radiation emitted by the disk will likewise be only weakly
dependent on magnetic topology.  It is possible that different magnetic
topologies could produce different vertical profiles of dissipation
(which these simulations are not designed to study), but the similarity
in magnetic pressure profiles (Fig.~\ref{angcor}) suggests that even
those differences are likely to be minor.

In regard to the accretion flow, we found only two dependences on field
topology worth noting: for fixed outer torus mass, the initial toroidal
field simulation gives a somewhat lower accretion rate \cite[a result
foreshadowed in][] {HK:2002}.  In addition, the magnetic field in the
plunging region is, in relative terms, somewhat stronger in the large
dipolar loop case than for quadrupolar or toroidal fields, so that
stresses near the ISCO are somewhat stronger when the field topology
is dipolar.  In other words, the short-wavelength behavior of the
MRI-driven MHD turbulence in the disk produces near-universal behavior,
independent of magnetic field structure.

Jets are quite different.  {\it Laminar} flow dynamics can, given the
right conditions (e.g., the right field topology), inflate field with no
pre-existing structure on scales larger than a disk scale-height into
a truly global field.  Simulations that begin with a large dipole loop
in the initial torus produce a strong Poynting-flux jet in the axial
funnel and a comparably strong matter-dominated funnel-wall outflow.
The inner field lines of the dipolar loop move quickly inward to fill the
funnel, and remain there until the opposite side of the loop is accreted,
an event which may not happen until quite a long time later. Thus,
this case mimics a situation in which net magnetic flux has been trapped
against the black hole event horizon by previous accretion.  When this
trapped flux creates a coherent poloidal field in the funnel, the black
hole's spin forces nearby field lines into rotation and drives a large
outward-directed Poynting flux.

This sequence of events is not as easily achieved with any
of the other field geometries we explored.   Unlike the dipole
configuration, the quadrupole
features a pair of current sheets located above and below the equator.
These current sheets are subject to instabilities that can lead to
reconnection in the magnetic field and loss of extended field coherence.
Even when the quadrupole is successful in establishing a funnel field,
the jet may live only briefly because fluctuations in the accretion
flow can generate reconnection events.  By the same token, quadrupolar
field can also restore field to an empty funnel.   Consequently, any
jet based on quadrupolar field tends to be unstable and episodic, with
the timescale and amplitude of changes set both by details of the field
geometry and the rate of reconnection.  Unfortunately, it is very hard
to obtain any information about the former in real systems, and in our
simulations the latter is controlled by numerical effects and
resolution, so it is impossible for us to make predictive statements
about the timescale or amplitude of jet variability in this case.

All of the models reported on above used a black hole with a spin $a/M
= 0.9$.  To test our conclusions against other black hole spins we
also ran two three-dimensional initially quadrupolar simulations
with black hole spins $a/M=0.0$ and, $a/M =0.998$.  The effect of a wide
range of black hole spins (including a counter-rotating hole) on dipolar
models was studied by \cite{Hawley:2006}.  No Poynting
flux jet is present for a Schwarzschild hole regardless of the initial
field topology.  The energy to drive the Poynting flux originates with
the spin of the hole.  However, the funnel is still filled with radial
field, which is considerably stronger in the dipole case compared to
the quadrupole.  On the other hand, the funnel field strength increases
with black hole spin for a given field topology.
For the rapidly-spinning hole, the Poynting flux jet is stronger than
in the $a/M = 0.9$ simulations for both topologies, but as we have discussed,
it is considerably weaker when the initial field is quadrupolar than when
it is dipolar.  Specifically, the ratio between the time-averaged Poynting
luminosity with dipolar field to the Poynting luminosity with quadrupolar
field is $\sim 350$ for $a/M = 0.998$ and $\sim
335$ for $a/M = 0.9$.  Thus, the accretion dynamics control what sort of
field is brought to the vicinity of the black hole and fill the funnel,
and the black hole spin determines how energetic the resulting jet is.

The purely toroidal field topology does not generate any
jet at all.  The only way for toroidal field to create any poloidal field
is by initiating non-axisymmetric perturbations of alternating sign.
The lengthscales of these perturbations must be relatively short, and
the portions of opposite sign must balance in magnitude.  Fields of this
sort can do very little to fill the funnel region because they are even
more subject to reconnection than is the quadrupolar field.

When the initial field is a single dipolar loop that wraps around the
initial pressure maximum of the simulation, all the vertical field brought
down to the funnel by the initial accretion has the same direction.
In real disks containing closed dipolar loops, oppositely directed
vertical field will, sooner or later, arrive in the inner disk, and
it may cancel the funnel field.  To investigate this possibility,
we ran a high-resolution axisymmetric simulation that began with
four narrow dipolar loops within the initial torus.  Its subsequent evolution
qualitatively supports this conjecture.  The innermost loop produces a
funnel field as in the single dipole loop model.  When the second loop,
oriented oppositely from the first one, descends, it displaces the field
established by the first loop.  The funnel field established by the first
loop reconnects, detaches from the event horizon, and is rapidly expelled
from the grid.  The second loop then expands poloidally and reestablishes
the funnel field.  In other words, the accretion of dipolar field
loops leads to jet creation, but its structure is highly-variable:
the narrower the dipolar loops present in the accretion flow, the more
variability we can expect in the jet.

When a simulation is initialized with (possibly multiple) dipolar field
loop(s) that cross the equatorial plane, the strength of the resulting
Poynting-flux jet depends on the strength of the magnetic field within
the disk itself and on the spin of the hole.  Poynting-flux jets that
are formed from quadrupolar fields are generically much weaker than
their dipolar counterparts, even when magnetic field strength
in the disk body is taken into account.  How much weaker depends on
other factors as well, including disk thickness, location of the inner
edge of the initial torus, and numerical resolution.  In the limited
range of numerical experiments reported here, we have seen suppression
factors ranging from several to several hundred.

Real accretion flows are likely to have a mixture, possibly time-varying,
of all these field topologies.  Real jets from black holes will exhibit
behavior combining the sorts of events we have seen in this paper:
stability over longer timescales when the field is mostly made of
large dipolar
loops, greater variability and occasional jet quenching when the field
has a larger contribution from higher multipoles, jet elimination when
the field is mostly toroidal.  It is possible that in addition
to the sort of coherent funnel field spontaneously created by large
closed-loop poloidal structures, there is a large-scale field imposed
by external boundary conditions.  We plan to investigate the impact of
such an externally-imposed field in future work.

This study is only one small step in the process of exploring
the mechanisms to create large scale field configurations within
accretion flows.  Here the large scale field formation process we have
observed results from an interaction between poloidal field components
in the accretion flow and the boundary represented by the black hole.
The mechanism operates in axisymmetry as well as in three dimensions, so
it is not a dynamo if that word is defined as in the antidynamo theorem.
Many authors have conjectured that an inverse cascade within the three
dimensional turbulent disk itself might generate large scale fields; we
see no evidence in these simulations of any disk field with structure on
scales more than a few scale-heights, but we caution that the resolution
limitations of three dimensional global simulations make this observation
less than definitive.

The results we have reported also stimulate speculations about whether
the contrasts in jet phenomenology from one object, or class of objects,
to another is due in part to differences in the mixture of different
field topologies, or the interaction of different topologies in the flow.
For example, could the jet outbursts sometimes seen in Galactic black
hole binaries when they cross the ``jet line" \citep{RM:2006} be due to
rapid field annihilation and loop expulsion of the sort observed in the
multiple dipolar loop simulation?  Perhaps the radio-weakness of most AGN
\cite[see e.g.][]{White:2007} is due to the predominance of small-scale
multipolar field in their accretion flows as a result of accreting from a
turbulent interstellar medium?  As we learn more about how it acts, we may
find that magnetic topology can, in fact, be inferred from phenomenology.

\acknowledgements{ This work was supported by NSF grant PHY-0205155 and
NASA grant NNG04GK77G (JFH), and by NSF grant AST-0507455 (JHK). KB thanks
Sean Matt for useful discussions. We acknowledge Jean-Pierre De~Villiers
for continuing collaboration in the development of the algorithms used
in the GRMHD code.  The simulations were carried out on the DataStar
system at SDSC.}

\begin{deluxetable}{cccccccc}
\tablecolumns{9}
\tablewidth{0pc}
\tablecaption{Coronal Diagnostics}
\tablehead{\colhead{Model}          &
\colhead{$\langle M_\mathrm{c} \rangle_{V} / \langle M_\mathrm{tot} \rangle_{V}$} &
\colhead{$\langle \beta_{\mathrm{c}} \rangle_{V}$} &
\colhead{$\langle E^\mathrm{therm}_{\mathrm{c}} \rangle_{V} / \langle M_\mathrm{c} \rangle_{V}$} &
\colhead{$\langle E^\mathrm{mag}_{\mathrm{c}} \rangle_{V}  / \langle M_\mathrm{c}\rangle_{V}$} &
\colhead{$\langle\mathrm{Vol_{c}}\rangle_{V} / \langle\mathrm{Vol_{tot}}\rangle_{V}$}
}
\startdata
KDPg & 0.055 &  3.6 & $2.9\times 10^{-3}$&$1.1\times 10^{-3}$& 0.48 \\
QDPa & 0.050 &  2.1 & $2.7\times 10^{-3}$&$1.8\times 10^{-3}$& 0.59 \\
TDPa & 0.054 &  3.4 & $2.5\times 10^{-3}$&$1.0\times 10^{-3}$& 0.61
\enddata
\label{corona}
\tablecomments{Subscript ``c'' denotes volume integral over the coronal region (as defined in the text), whereas subscript ``tot'' denotes volume integral over the total simulation region.}
\end{deluxetable}

\clearpage

\end{document}